\documentclass[11pt]{article}
\usepackage{amsmath}
\usepackage{amsfonts}   
 \usepackage{amssymb} 
 \usepackage{fullpage}
 
 \usepackage{maplestd2e}

\DefineParaStyle{Maple Heading 1}
\DefineParaStyle{Maple Text Output}
\DefineParaStyle{Maple Dash Item}
\DefineParaStyle{Maple Bullet Item}
\DefineParaStyle{Maple Normal}
\DefineParaStyle{Maple Heading 4}
\DefineParaStyle{Maple Heading 3}
\DefineParaStyle{Maple Heading 2}
\DefineParaStyle{Maple Warning}
\DefineParaStyle{Maple Title}
\DefineParaStyle{Maple Error}
\DefineCharStyle{Maple Hyperlink}
\DefineCharStyle{Maple 2D Math}
\DefineCharStyle{Maple Maple Input}
\DefineCharStyle{Maple 2D Output}
\DefineCharStyle{Maple 2D Input}

\newcommand{\zetab}{\overline\zeta}
\newcommand{\Maple}{{\sl Maple }}
\newcommand{\DG}{{\sl DifferentialGeometry}}
\newcommand{\next}{\bigskip\noindent}
\newcommand{\caret}{\^{}}
\newcommand{\half}{\frac{1}{2}}
\newcommand{\proc}{\tt}
\newcommand{\DiffM}{{\it Diff(M)}}
\numberwithin{equation}{section}

\title{New Symbolic Tools for Differential Geometry,  Gravitation, and Field Theory}
\author{I.~M.~Anderson\\
{\it Department of Mathematics and Statistics, Utah State University, Logan, Utah 84322, USA}\\
\\
 C.~G.~Torre\\
{\it Department of Physics, Utah State University, Logan, Utah 84322, USA}}

   \date{March 7, 2011}

\begin{document}
 
  \maketitle

 \abstract{\DG\ is a Maple software package which symbolically performs fundamental operations of calculus on manifolds, differential geometry, tensor calculus, Lie algebras, Lie groups, transformation groups, jet spaces, and the variational calculus. These capabilities, combined with dramatic recent improvements in symbolic approaches to solving algebraic and differential equations, have allowed for development of powerful new tools for solving research problems in gravitation and field theory.  The purpose of this paper is to describe some of these new tools and present some advanced applications involving:  Killing vector fields and isometry groups, Killing tensors and other tensorial invariants, algebraic classification of curvature, and symmetry reduction of field equations.}
 
 \vfill\eject
 
 \section{Introduction}
 
 \DG\ is a Maple software package which symbolically performs fundamental operations of calculus on manifolds, differential geometry, tensor calculus, Lie algebras, Lie groups, transformation groups, jet spaces, and the variational calculus. These capabilities, combined with dramatic recent improvements in symbolic approaches to solving algebraic and differential equations, have allowed for development of powerful new tools for solving research problems in gravitation and field theory.  The purpose of this paper is to describe some of these new tools and present some advanced applications. 

In the next section we show how to symbolically define a chart on a manifold, define tensor fields on the manifold and perform various routine computations.  Such computations feature in all existing symbolic approaches to the present subject (see, {\it e.g.}, \cite{ACGR}). We then focus on a number of applications that take advantage of new mathematical functionalities which, to the best of our knowledge, have not previously been available in an integrated software package. 

In \S3 we consider the problem of finding Killing vector fields and determining the structure of the associated isometry group.  As an example we show that the G\"odel spacetime is a reductive homogeneous space. Also demonstrated in \S3 is how to use \DG\ to perform case-splitting analysis, which isolates exceptional values of functions and parameters. This latter point is illustrated with a family of explicit vacuum metrics in the Kundt class which generically admit no isometries.

In \S4 we consider the various issues associated with algebraic classification of curvature tensors.  We again focus on the ability to perform case splitting when solving equations symbolically to analyze a family of hypersurface homogeneous spacetimes found in \cite{Stephani2003}. With one minor correction we confirm the results there and we show that the family of metrics includes some additional, new Petrov types at exceptional values of parameters appearing in the metric.  Each Petrov type determines a family of adapted null tetrads, a set of principal null directions, and a factorization of the Weyl spinor,  each of which can be obtained symbolically.  This we illustrate with the G\"odel spacetime.

In \S5 we discuss how the vector bundle functionality of \DG\ allows one to symbolically implement the formalism of spinor fields. Currently, 
\DG\ has a full implementation of the most important case:  the 2-component spinor formalism on a four-dimensional pseudo-Riemannian manifold. We illustrate this by finding the principal spinors, which factorize the Weyl spinor, for the G\"odel spacetime.  The algorithms for computing adapted null tetrads and principal null directions make use of these principal spinors. 

  In \S6 we focus on various tensorial invariants of (pseudo-)Riemannian manifolds such as homothetic vector fields, Killing tensors and Killing-Yano tensors.  In particular, (i) we compute the homothety group for plane wave spacetimes; (ii) we show how to compute the rank-2 Killing-Yano tensor and associated Killing tensor for Kerr spacetime; and (iii) we show that the G\"odel spacetime has no non-trivial Killing tensors of rank-2 and rank-3.  
  
 Finally, in \S7 we symbolically analyze the process of reduction of field equations by a prescribed group of symmetries in the context of gravitational plane waves.   In particular we show how \DG\ can be used to find the most general group-invariant metric and to find the residual symmetry group admitted by the reduced field equations.

In \DG\ all tensor fields, spinor fields, connections, and so forth are built and displayed using index-free notation as is standard in the mathematical literature. \DG\ runs as a package in {\sl Maple} (see, {\it e.g.,} \cite{Heck2003}) and the syntax of all commands and output which we describe reflects this.\footnote{Note, in particular, that a command terminated with a colon will suppress displaying the results of the command, while a command terminated with a semi-colon will display the results of the command.} We have taken a few liberties with the output of some of the commands to make it more readable in this venue. See the Appendix for a complete list of all the available commands.  \DG\  first appeared in Maple 11; much of the capabilities described here appear in Maple 14 and Maple 15. Full documentation and tutorials can also be found in these releases. We note that the underlying code for all \DG\ commands can be explicitly displayed so that, if desired, one can have complete control over the details of all computations.   

It is a pleasure to thank Edgardo Cheb-Terrab, first for proposing the development of general relativity tools within the framework of \DG, and second, for his time and efforts to integrate this work into the Maple distributed library of packages. Support from the National Science Foundation (DMS-0713830) and Maplesoft is gratefully acknowledged.

 \section{Manifolds, tensor fields, frames}
 
 When analyzing a (pseudo-)Riemannian manifold using \DG\   the first order of business is to define a chart on the manifold where computations will take place and to specify a metric (or some other element of structure).  Defining the chart in \DG\ simply amounts to specifying the names of coordinates which will be used and specifying a name for the chart. (Other charts and coordinates can be defined subsequently as desired.)
The command for defining a chart is $\tt DGsetup$. Here we define a chart named $M$ with coordinates $(t,r,\theta,\phi)$.
 
 \next
 {\proc > DGsetup([t, r, theta, phi], M):}

\next 
This specifies the dimension of the manifold, it protects the coordinate labels and it defines coordinate basis vector fields and forms.  

All tensor operations are supported; there are commands for scalar multiplication, tensor product, tensor addition, contraction of indices, symmetrization of indices, and so forth. The command {\tt evalDG} can be used to allow a streamlined syntax for some of these operations. In particular, within {\tt evalDG} one can build tensor fields using the asterisk for scalar multiplication, using $\pm$ for tensor addition/subtraction, using {\tt \&t} for tensor product, using {\tt \&w} for wedge product, and using {\tt \&s} for symmetric tensor product.

One can now specify fields on the manifold $M$. For example, one often wants to begin by specifying a metric in the coordinate basis. We do this as follows using the the Reissner-Nordstr\"om metric with mass $M$ and charge $Q$ as an example. For convenience, we first define some standard notation:

\next
{\proc > f := 1 - 2*M/r + Q\caret{2}/r\caret{2}:

\next
 {\proc > dOmega := evalDG(dtheta \&t dtheta + sin(theta)\caret{2}*dphi \&t dphi}):}

\next
The metric is now defined by

\next
{\proc > g := evalDG(-f*dt \&t dt + 1/f*dr \&t dr 
+ r\caret{2}*dOmega);}

\begin{equation}
\label{ }
g := (-1 +  {2M\over r} - \frac{Q^{2}}{r^{2}}) dt\ dt + \frac{r^{2}}{r^{2} - {2M}{r} +Q^{2}} dr\ dr
+ r^{2} (d\theta\ d\theta + \sin(\theta)^{2} d\phi \ d\phi)\nonumber
\end{equation}

\next 
Note that the tensor product is implicit in the \Maple output.
Next we define the Reissner-Nordstr\"om electromagnetic field $F$ as a spacetime 2-form:

\next
{\proc > F := evalDG(sqrt(2)*Q/r\caret2*dt \&w dr);}

\begin{equation*}
F := \frac{\sqrt{2}\, Q}{r^2}\, dt \wedge dr
\end{equation*}

A number of frequently used tensor fields are pre-defined for convenience. For example, energy momentum tensors can be computed once the type of field and relevant geometric data are supplied. Here we compute the energy-momentum tensor of the Reissner-Nordstr\"om electromagnetic field:

\next
{\proc > T := EnergyMomentumTensor(``Electromagnetic'', g, F);}

\begin{equation*}
T :=  \frac{Q^{2}}{r^{2}(r^{2} - {2M}{r} +Q^{2})} D_{t} D_{t} - \frac{Q^{2}(r^{2} - {2M}{r} +Q^{2})}{r^{6}} D_{r} D_{r} + \frac{Q^{2}}{r^6} D_{\theta} D_{\theta} + \frac{Q^{2}}{r^6\sin(\theta)^{2}} D_{\phi} D_{\phi}
\end{equation*}
Note that $T$ has been defined as a tensor of type $\left(2\atop0\right)$; the quantities $(D_{t}, D_{r} D_{\theta}, D_{\phi})$ are the coordinate basis vector fields.
Here we  verify that $g$ and $F$ determine a solution to the Einstein-Maxwell equations. First we check the Einstein equations:

\next
{\proc > G:= EinsteinTensor(g):}

\next
{\proc > evalDG(G - T);}

\begin{equation*}
0\, D_{t} D_{t} 
\end{equation*}

\next
Next we check the source-free Maxwell equations by verifying $\nabla\cdot F=0$ and $dF=0$. This can be done using the {\tt CovariantDerivative} command and the various tensor operations. Again, for convenience, these computations are pre-defined. $\nabla\cdot F$ and $dF$ are given by, respectively,

\next
{\proc > MatterFieldEquations(``Electromagnetic'', g, F);}

\begin{equation*}
0\, D_{t},\quad 0\, dt \wedge dr\wedge d\theta
\end{equation*}

\next

It is also possible to define fields relative to a given anholonomic frame.  This is often the most effective setting in which to perform complex computations.   We illustrate this with the Reissner-Nordstr\"om spacetime.  

We begin by defining an orthonormal tetrad $OT$ as a list of vector fields. 

\next
{\proc > OT := evalDG([1/sqrt(f) * D\_{t}, sqrt(f) * D\_r, D\_theta/r, D\_phi/(r*sin(theta))]);}

\begin{equation}
\label{ }
OT := [\frac{r}{\sqrt{r^{2} -  {2Mr} +Q^{2}}} D_t, \frac{\sqrt{r^{2} -  2Mr +Q^{2}}}{r}D_r, \frac{1}{r} D_{\theta}, \frac{1}{r\sin(\theta)}D_{\phi}]
\nonumber
\end{equation}
We can verify this is in fact an orthonormal tetrad by computing the 10 scalar products between these vector fields using the metric $g$. A shortcut is the {\tt TensorInnerProduct} command which will take any two (lists of) tensors of the same type and contract all indices with a specified metric. Using the list $OT$ the output is the matrix of scalar products:

\next
{\proc > TensorInnerProduct(g, OT, OT);}

$$
\begin{pmatrix}
-1 &0 &0 &0\\ 0 &1 &0 &0\\ 0 &0 &1 &0\\ 0 &0 &0 &1
\end{pmatrix}
$$

\next
When working in an orthonormal frame the structure equations of the frame, {\it i.e.,} the commutation relations of the frame vector fields, determine all the geometric properties of the spacetime.  These structure equations are computed and stored for future use using the commands {\tt FrameData} and {\tt DGsetup}, respectively:

\next
{\proc > FD := FrameData(OT, O);}

\begin{align*}
\label{ }
FD := &\Big[[E1, E2] = -\frac{M  r - Q^{2}}{r^{2} \sqrt{r^{2}-2 M r + Q^{2}}} E1, \ 
[E2, E3] = -\frac{\sqrt{r^{2}-2 M r + Q^{2}} }{r^{2}} E3,\ \nonumber\\ 
&[E2, E4] = -\frac{\sqrt{r^{2}-2 M r + Q^{2}} }{r^{2}} E4,\  [E3, E4] = -\frac{\cos(\theta) }{r \sin(\theta)}E4\Big]
\end{align*}

\next
{\proc > DGsetup(FD):}

\next
The (default) labeling for the basis vectors is $(E{1}, E{2},E{3}, E{4})$, the (default) labeling for the dual basis is $(\Theta{1},\dots,\Theta{4})$. The variable $FD$ contains all their commutators as well as the name ($O$) chosen for the anholonomic frame. 

One can now perform all computations in the given orthonormal frame.  Here we define the metric in this orthonormal frame:

\next
{\proc > eta := evalDG(- Theta1 \&t Theta1 + Theta2 \&t Theta2 

+ Theta3 \&t Theta3 + Theta4 \&t Theta4);}

\begin{equation*}
\label{ }
\eta := -\Theta1 \Theta1 +\Theta2 \Theta2 + \Theta3 \Theta3 + \Theta4 \Theta4
\end{equation*}

\next
The Ricci tensor takes the form:

\next
{\proc > Ricci := RicciTensor(eta);}

\begin{equation*}
Ricci := \frac{Q^{2}}{r^{2}}\, \Theta1\, \Theta1 -\frac{Q^{2}}{r^{2}} \,\Theta2\, \Theta2+\frac{Q^{2}}{r^{2}}\, \Theta3\, \Theta3+\frac{Q^{2}}{r^{2}} \,\Theta4\, \Theta4
\end{equation*}

 \section{Isometries and Killing vector fields}

Let $g$ be a metric on a manifold $M$ and $G$ a Lie group.
   A {\it group action} is a smooth mapping $\mu\colon G\times M \to M$.  The {\it isometry group} of the metric $g$ is a group action which preserves the metric, {\it i.e.}, for each $h\in G$,  $\mu^{*}_{h}g = g$.  
The infinitesimal generators of the isometry group are the Killing vector fields $V$, which satisfy an over-determined system of linear PDEs, the Killing equations:
\begin{equation}
\label{Killingeq}
L_{V} g_{ab} = \nabla_{a} V_{b} + \nabla_{b} V_{a} = 0.
\end{equation}
Using the vector field commutator as the Lie bracket, the vector space of Killing vector fields forms a Lie algebra $\mathfrak g$ isomorphic to the Lie algebra of $G$.   
In this section we show how to symbolically determine Killing vector fields and analyze the corresponding isometry group.

 \subsection{Computation of Killing vector fields}
 
 Our first illustration of computation of Killing vector fields uses the G\"odel metric, which is a homogeneous perfect fluid solution of the Einstein equations. Expressed in coordinates $(t, x, y, z)$ it takes the form
 \begin{equation}
\label{Godel}
g = - (dt + e^{x} dz) \otimes (dt + e^{x} dz) + dx \otimes dx + dy \otimes dy + \half e^{2x} dz \otimes dz.
\end{equation}
The metric can be defined within \DG\  via

\next
{\proc > DGsetup([t, x, y, z], M):}

\next
{\proc > omega := evalDG(dt + exp(x)*dz):}

\next
 {\proc >  g := evalDG(- omega \&t  omega + dx \&t dx + dy \&t dy + (1/2)*exp(2*x)*dz \&t dz)}:
 \bigskip
 
 The {\proc KillingVector} command computes the Killing equations (\ref{Killingeq}) of the metric $g$ and passes the resulting partial differential equations to the \Maple differential equation routines. Applied to the G\"odel metric, the output is a list of 5 Killing vector fields which form a basis for the vector space of solutions to (\ref{Killingeq}):
 
 \next
 {\proc > KV := KillingVectors(g);}
 
 \next
 \begin{equation}
\label{ }
KV := [D_{z}, D_{t}, D_{x} - z D_{z}, - 4 e^{-x} D_{t} + 2 z D_{x} +(- z^{2} + 2 e^{-2x} ) D_{z}, D_{y}]\nonumber
\end{equation}
 One can verify that these vector fields are indeed Killing vector fields by taking the Lie derivative of the metric with the {\proc LieDerivative} command. 

\next
{\proc > LieDerivative(KV, g);}
\next
\begin{equation}
\label{ }
[0\, dt\, dt, 0\,dt\, dt, 0\,dt\, dt, 0\,dt\, dt, 0\,dt \,dt]
\nonumber
\end{equation}
 where this list of five vanishing tensors corresponds to the Lie derivative of the metric with respect to each vector fields in the list $KV$.
 
As an independent confirmation that all solutions to the Killing equations have been found, we can take advantage of the fact that the abstract Lie algebra $\mathfrak g$ of the isometry group can be determined purely in terms of geometric data at a single point (see, {\it e.g.,} \cite{Ashtekar1978}).  In particular, the dimension of the space of  Killing vector fields and their commutators can be computed without solving the Killing equations. These facts are utilized in the {\proc IsometryAlgebra} command, which takes as arguments the metric, a basepoint, and a label for the resulting Lie algebra. The output consists of a list specifying the bracket relations of a basis for the Lie algebra of Killing vector fields.
For simplicity, in our G\"odel metric example we take the basepoint to be $(t=0, x=0, y=0, z=0)$. The Lie algebra  of the isometry group is then given as follows.

\next
{\proc IsometryAlgebra([g], [t=0, x=x0, y=0, z=0], Godel, [1]);}

\next
\begin{align}
\label{ }
&[e1, e2] = e1-e4-\frac{1}{2}e5,\quad [e1, e4] = \frac{1}{2}e2,\quad  [e2, e4] = -e1+2e4+e5,\nonumber\\ 
&[e2, e5] = 2e1-2e4-e5,\quad [e4, e5] = e2
\end{align}
This result shows that there are indeed 5 Killing vector fields, corresponding to a Lie algebra admitting a basis $(e_{1}, e_{2}, \dots, e_{5})$ with the indicated brackets. 
%
%
 
 \subsection{Properties of isometry algebras}

Given the Killing vector fields admitted by a metric, the isometry group they generate can be characterized using a number of techniques available in the \DG\  package. Such a characterization is useful, of course, for better understanding the properties of the geometry defined by the metric. But characterization of the isometry group is also important for analyzing the equivalence problem for metrics, since the  isometry group is a diffeomorphism invariant of the metric.  Characterization of the isometry group naturally splits into two parts: characterization of the abstract group $G$ and characterization of the group action $\mu$.  We illustrate this with the G\"odel spacetime.

We can characterize the abstract group $G$ by determining its Lie algebra and then computing its Levi decomposition. Recall that every Lie algebra $\mathfrak g$ can be decomposed into  a semi-direct sum of a solvable and a semi-simple Lie algebra, respectively. The solvable ideal is unique as is the quotient of $\mathfrak g$ by the ideal.  Given the Killing vectors of the G\"odel metric, 

\next
{\proc > KV := KillingVectors(g);}

\begin{equation}
\label{ }
KV := [D_z, D_t, D_x-z D_z, -4 e^{-x} D_t+2 z D_x+(-z^2+2 e^{-2 x}) D_z, D_y]
\nonumber
\end{equation}
we define the Lie algebra as a data structure $LD$, which stores  the vector space and  brackets which define the Lie algebra $\mathfrak g$. 

\next
{\proc > LD := LieAlgebraData(KV, A);}

\next
{\proc > DGsetup(LD)}:

\begin{equation}
\label{ }
LD := [[e1, e3] = -e1, \quad [e1, e4] = 2\,e3, \quad [e3, e4] = -e4]
\nonumber
\end{equation}

\noindent
The Lie algebra is now defined internally with a basis $(e_{1}, e_{2},\dots,e_{5})$ corresponding to the 5 vector fields in $KV$.
We next compute the Levi decomposition of $\mathfrak g$

\next
{\proc > LeviDecomposition(A);}

\begin{equation}
\label{ }
[[e2, e5], [e1, e3, e4]]
\nonumber
\end{equation}

\next
The output displays a basis for the solvable ideal, $[e2, e5]$, and a basis for a complementary semi-simple sub-algebra, $[e1, e3, e4]$, in terms of the original basis $[e1,\dots, e5]$ of $\mathfrak g$.  It is easy to see by inspection of the Lie brackets in $\mathfrak g$ that the Levi decomposition is in this case actually a direct sum decomposition of the solvable algebra and $sl(2,R)$. Alternatively,  the \DG\  command {\tt Decompose} could have been used to find this direct sum decomposition.  

Let us now examine some features of the group action $\mu$. Two salient properties of any group action are the dimension of the orbits and the isotropy subgroups.  The dimension of the orbits are easily seen by inspection to be four in the G\"odel example --- the Killing vectors span the tangent space at each point, so the spacetime is a homogeneous space.  In more complicated examples the dimension and signature of the orbits can be computed using the \DG\ command {\tt SubspaceType}. Let us focus on the isotropy subgroups.  

Recall that the isotropy group of a point $p\in M$ is the subgroup $G_{p}\subset G$ such that $\mu_{h}(p) = p, \ \forall h\in G_{p}$.  The isotropy subalgebra $\mathfrak g_{p} \subset \mathfrak g $ of a point $p\in M$ is the subalgebra of Killing vector fields which vanish at $p$.  This subalgebra can be determined from a given algebra of vector fields with the {\tt IsotropySubalgebra }command. Here we compute the isotropy subalgebra of the origin for the Killing vector fields $KV$ of the G\"odel metric and express it in terms of the algebra $A$ defined above:

\next
{\proc > Iso := IsotropySubalgebra(KV, [t=0, x=0, y=0, z=0], output=["Vector", A]);}

\begin{equation}
\label{ }
Iso := [(2e^{-x}-2)D_t-zD_x+((1/2)z^2-e^{-2x}+1)D_z], \quad [e1-2\,e2- \frac{1}{2}\,e4]
\nonumber
\end{equation}

\next
Thus there is a one dimensional isotropy subgroup of the origin whose infinitesimal generator is the Killing vector field
\begin{equation}
\label{ }
V=2(e^{-x}-1)\partial_t-z\partial_x+(\half z^2-e^{-2x}+1)\partial_z = e1-2\,e2- \frac{1}{2}\,e4.
\end{equation}

Since $G$ acts by isometry, the isotropy subgroup of a point $p\in M$ preserves the inner product defined at $T_{p}M$ by the Lorentz-signature metric, and hence can be identified with a subgroup of the Lorentz group.  Infinitesimally, the isotropy subalgebra is a subalgebra of the Lorentz algebra consisting of combinations of boosts, spatial rotations and null rotations. All such subalgebras have been classified up to conjugation \cite{Patera1975} (for four dimensional spacetimes).   Given a set of Killing vector fields and a point $p\in M$ the {\tt IsotropyType} command computes this classification of the isotropy subalgebra of $p$. For the Killing vector fields $KV$ of the G\"odel metric the isotropy subalgebra of the origin is  one dimensional and classified by

\next
{\proc > IsotropyType(KV, [[t=0, x=0, y=0, z=0]);}

\begin{equation}
\label{ }
{\rm F12}
\end{equation}

\next
Since the G\"odel spacetime is homogeneous, this is in fact the isotropy type of any point.
In the classification scheme of \cite{Patera1975}, the subalgebra $F_{12}$ corresponds to the generator of a spatial rotation. Thus the G\"odel spacetime is locally rotationally symmetric.

Another important property of the group action, which characterizes how the isotropy group sits as a subgroup of $G$, is whether its Lie algebra $\mathfrak{h}\equiv\mathfrak g_p$ admits a complement $\mathfrak m$ in $\mathfrak g$ such that $[{\mathfrak h}, {\mathfrak m}] \subset {\mathfrak m}$. If so, the vector space decomposition ${\mathfrak g} = {\mathfrak h} + {\mathfrak m}$ is called a {\it reductive} decomposition of ${\mathfrak g}$ and the group orbit is a {\it reductive homogeneous space} \cite{Kobayashi1969}.  We now show symbolically that the isometry algebra of the G\"odel spacetime does admit a reductive decomposition.  We have already symbolically defined the isometry algebra $A$ and its isotropy subalgebra $Iso$; we now compute the general form of its complement:

\next
{\proc > H := Iso[2];}

\begin{equation*}
H := e_{1} - 4 e_{2} + e_{4}
\end{equation*}

\next
{\proc > M := ComplementaryBasis(H, [e1, e2, e3, e4, e5], t);}

\begin{align*}
M := &[(t_1+1)e_1-2t_1e_2 - \frac{1}{2}t_1e_4, t_2e_1+(-2t_2+1)e_2-\frac{1}{2}t_2e_4, \\
&t_3e_1-2t_3e_2+e_3-\frac{1}{2}t_3e_4, t_4e_1-2t_4e_2-\frac{1}{2}t_4e_4+e_5], \{t_1, t_2, t_3, t_4\}
\end{align*}

The result indicates that there is a 4 parameter family of complementary subspaces (parametrized by $t_1,\dots,t_4$) and gives a basis for the subspaces. The {\tt Query} command can now determine if any of these complements defines a reductive decomposition and, if so, display a basis for the reductive subspace $\mathfrak{m}$:

\next
{\proc > Query(H, M, "ReductivePair")[-1][1][2];}

\begin{align*}
&[( t_2+1) e_1+(-2 t_2+1) e_2+(-\frac{1}{2}t_2+\frac{1}{4}) e_4, \quad t_2 e_1+(-2 t_2+1) e_2-\frac{1}{2}t_2 e_4,\quad e_3, \\
&t_4 e_1-2 t_4 e_2-\frac{1}{2}t_4 e_4+e_5]
\end{align*}
Thus there is in fact a two parameter family of reductive decompositions of $\mathfrak g$ relative to $\mathfrak h$. 
 
 \subsection{Case splitting}
 
 Here we present an example of case splitting in the analysis of Killing vector fields.   The metric we shall consider comes from the class of spacetimes admitting a non-diverging, shear-free, null geodesic congruence \cite{Kundt1961, Stephani2003, DMorse}.  The metric $g$ is expressed in coordinates $(u,v, \zeta, \overline\zeta)$ as
\begin{equation}
\label{g}
g = d\zeta \odot d\zetab - du \odot (dv + W d\zeta + \overline W d\zetab + H du),
\end{equation}
where $\odot$ is the symmetric tensor product and
\begin{equation}
\label{ }
W = - {2v\over \zeta + \zetab},\quad H = (\zeta+\zetab)f(u)(e^{\zeta}+e^{\zetab})- {v^{2}\over (\zeta + \zetab)^{2}}
\end{equation}
A bar on a quantity denotes complex conjugation.

It is readily verified using \DG\  functionality that the vector field $\partial_{v}$ determines a non-diverging, shearfeee, null geodesic congruence and, furthermore, that this is a vacuum metric of Petrov type $N$ in any open region where $f(u)\neq 0$. The spacetime is flat wherever $f(u)=0$; for simplicity we will suppose $f(u)\neq0$ in what follows.   
 
The spacetime can be defined within \DG\  via

 \bigskip
 \noindent
 {\proc > DGsetup([u,v,zeta, zetab], M):}
 
 \bigskip
 \noindent
 {\proc > xi := evalDG(dv + W*dzeta + W*dzetab + H*du):}

\next
{\proc > W := - 2*v/(zeta + zetab):}

\next
{\proc > H := (zeta + zetab)*  f(u)* (exp(zeta) + exp(zetab)) - v\caret2/(zeta + zetab)\caret2:}

\next
{\proc > g := evalDG(dzeta \&s dzetab - du \&s xi):}
 
\bigskip

We first look for solutions to the Killing equations making no specific choices for the function $f(u)$, in which case the PDE solvers treat $f(u)$ as generic. The result is that there are no non-trivial solutions:

\next
{\proc > KillingVectors(g);}

\smallskip\noindent
\centerline{[\ ]}

\next
Thus (\ref{g}) provides an apparently rare instance of an {\it explicit family of vacuum solutions to the Einstein equations which admit no continuous symmetries}.  This result can be independently verified using the analysis provided by the {\tt IsometryAlgebra} command, as discussed previously.

For generic choices of $f(u)$ there are no continuous isometries. However, it is possible that specific choices of $f(u)$ allow for the existence of Killing vector fields. This possibility is analyzed by adding keyword arguments ``parameters'' and ``auxiliaryequations''  to the {\tt KillingVector} command.  These keywords respectively identify the parameters and/or functions which are to be investigated for exceptional values, and impose any desired restrictions on the parameter/functions.  The syntax of the command is

\next
{\proc > KV := KillingVectors(g, parameters=[f(u)], auxiliaryequations = \{f(u) <> 0\});}

\next
The  ``auxiliaryequations'' keyword is used to eliminate the trivial case $f(u)=0$ from consideration.
The output of this command consists of a sequence of lists of Killing vector fields followed by a list of the corresponding conditions on $f(u)$ which allow them:

\begin{align}
\label{KVoutput}
KV  :=&[\ ], [(-2 C_1+\half u^2+C_2 u+ \half C_2^2) D_u
-(\half \zeta^2
+\zeta \zetab+v C_2+\half \zetab^2+v u) D_v], \nonumber\\
&[{f(u) = f(u)}, 
{f(u) = C_3/(4 C_1-u^2-2 C_2 u-C_2^2)^2}]\nonumber\\
\end{align}

\next
We thus obtain two classes of results. The first entry is an empty list which corresponds to the condition $f(u)=f(u)$; this recovers the result that for generic choices of $f$ there are no solutions to the Killing equations for this vacuum metric.   
The second entry is a three parameter family of Killing vector fields corresponding to a  three parameter family of  choices for $f(u)$, 
\begin{equation}
\label{f}
f(u) = {c\over a - b u - u^{2}}.
\end{equation}
Thus for each generic choice of $a$, $b$, $c$ in (\ref{f}) there is a single Killing vector field
\begin{equation}
\label{ }
( a - b u - u^{2}) \partial_{u} +\left[(\zeta + \zetab)^{2} - v(b + 2u)\right]\partial_{v}.
\end{equation}

Subsequent case splitting analysis of this 3 parameter family of metrics defined by (\ref{f}) reveals that the maximal one-dimensional isometry group exists for all values of the parameters except $c=0$, which corresponds to flat spacetime.

\section{Algebraic properties of curvature}
\label{Algebraic}

The algebraic classification of curvature --- ``Petrov type'' for classification of the Weyl tensor and ``Segre type'' for classification of the Ricci tensor --- is an extremely important tool for invariantly characterizing spacetimes.  The analysis needed to find the Petrov or Segre type is relatively straightforward, mathematically speaking,  but it can become quite involved computationally, so here is a place where computer algebra systems can be quite useful.  The algorithms used by \DG\  for algebraic classification of curvature are a refinement  of those devised in \cite{Acvevedo2006, Zakhary2004}.  The input to the algorithm is a null tetrad which the user may supply or which is constructed automatically by \DG. The principal difficulties which arise in determining the Petrov type symbolically include: (i) excessive computational times owing to a poor choice of null tetrad; (ii) inability to determine whether various invariants are equal or vanish (``zero recognition problem'' for scalars); and (iii) the possibility of complicated branching of results depending upon exceptional values of parameters or functions.     Finding a computationally effective null tetrad can be a bit of an art, but the range of possibilities can be explored by applying various Lorentz transformations via the {\tt NullTetradTransformation} commands. The zero recognition problem is a fundamental issue in symbolic analysis; in \DG\ this is ameliorated by allowing the user to control and inspect all aspects of the classification process. In particular, the {\tt infolevel} command allows the user to inspect each step of the classification algorithm, the {\tt Preferences} command allows the user to create custom-made simplification rules, and the {\tt output} option allows the user to examine the various algebraic and differential equations which arise in determining the algebraic classification.  Possible branching of the classification is handled using customizable case-splitting analysis options.
We illustrate some of this with the following computations of Petrov type.

\subsection{Petrov Type -- inspecting the algorithm}


Here we give an elementary example of a spacetime whose Petrov type is not constant and so a ``black box'' approach to computing the Petrov type is not appropriate.  We look inside the ``black box'' with the {\tt infolevel} command.

We define the spacetime:

\next
{\proc >  DGsetup([t, x, y, z], M):}

\next
{\proc > g := evalDG( (1/x\caret2)*dt \&t dt - (1/t\caret2)* dx \&t dx - dy \&t dy - dz \&t dz);}

\begin{equation}
g := \frac{1}{x^2} dt \, dt - \frac{1}{t^2} dx\, dx - dy\, dy - dz\, dz
\end{equation}

A simple call to PetrovType can be used to give the algebraic classification of the Weyl tensor at a generic spacetime point:

\next
{\proc > PetrovType(g);}

\begin{equation*}
``D"
\end{equation*}

\next
Details of the computation can be displayed using the {\tt infolevel} command.  To keep the expressions manageable, we compute the Petrov type at a point $x^\alpha=(t_0, x_0, y_0, z_0)$ and explicitly limit computations to the domain $t_0>0$ and $x_0>0$.

\next
{\proc > infolevel[PetrovType] := 2:}

\next
{\proc > PetrovType(g, [t0, x0, y0, z0])) assuming t0 > 0, x0 > 0;}

\next
{\sl The NP Weyl curvature scalars are:

   Phi[0]: 1/2*(-x0\caret4+t0\caret4)/t0\caret2/x0\caret2
   
   Phi[1]: 0
   
   Phi[2]: -1/6*(-x0\caret4+t0\caret4)/t0\caret2/x0\caret2
   
   Phi[3]: 0
   
   Phi[4]: 1/2*(-x0\caret4+t0\caret4)/t0\caret2/x0\caret2

Checking type O (Psi = 0):

   not type O

Test to see if the Weyl scalars are in Penrose-Rindler normalized form

\centerline{``D''}}

\next
As can be seen from this output, the algorithm first computes the Newman-Penrose Weyl curvature scalars, which are to be used to construct various scalar curvature invariants determining the Petrov type. The curvature scalars are first tested to see if they all vanish, in which case the Petrov type is O. It is clear from the scalar invariants displayed above that they are non-vanishing everywhere except on the null hypersurfaces $x=\pm t$. Away from these hypersurfaces, the curvature scalars are in a normal form for type D spacetimes and the algorithm terminates. By computing the Petrov type at an exceptional point, say $x=t =a$, we confirm that the spacetime is of type $O$ at such points:

\next
{\proc > PetrovType(g, [t=a, x=a, y=y0, z=z0]);}

\centerline{\it ``O''}

\next
\subsection{Petrov Type -- case splitting} 

The system of equations determining the Petrov type of a spacetime are amenable to case splitting analysis, which we now explore with a family of hypersurface homogeneous spacetimes  determined by a parameter $a$ and a function $f$ (see \cite{Stephani2003} \S 13.3.3):
\begin{align}
ds^{2} = &-2 \left(x^{a} du - \frac{dy}{(a+1)x}\right)\left[dt + a(t dx + dy)/x + f(t) \left( x^{a} du - \frac{dy}{(a+1)x}\right)\right]\nonumber\\
& + (dx^{2} + dy^{2} )
\frac{(t^{2}+1)}{2x^{2}}
 \label{1367}
\end{align}

\next
This metric can be defined as follows.

\next
{\proc > DGsetup([t, u, x, y], M):}

\next
{\proc > alpha := evalDG(x\caret a*du - 1/(a+1)/x*dy);}

\begin{equation}
\alpha := x^{a} du - \frac{1}{(a+1)x} dy
\end{equation}

\next
{\proc > beta := evalDG(dt + a/x*(t*dx + dy) + f(t)*alpha);}

\begin{equation}
\beta = dt + f(t) x^{a} du + \frac{at}{x}  dx +  (\frac{a}{x}  - \frac{f(t)}{(a+1)x})dy  
\end{equation}

\next
{\proc > g := evalDG(-alpha \&s beta + (t\caret2+1)/2/x\caret2*(dx \&t dx + dy \&t dy)):}

\next

Using a combination of the commands {\tt DGGramSchmidt} to construct an orthonormal tetrad, {\tt NullTetrad} to construct the corresponding null tetrad, and {\tt NullTetradTransformation} to simplify the result, the we obtain the following null tetrad:

\begin{align*}
\label{ }
NT:=&[D_t,- f(t) D_t+x^{-a} D_u,\\
&-\frac{a (I+t) }{\sqrt{t^2+1}}D_t+\frac{I x^{-a}}{\sqrt{t^2+1} (a+1)} D_u+\frac{x}{\sqrt{t^2+1}} D_x+\frac{I x}{\sqrt{t^2+1}} D_y,\\
&\frac{a (I-t)}{\sqrt{t^2+1}} D_t-\frac{I x^{-a} }{\sqrt{t^2+1} (a+1)}D_u+\frac{x}{\sqrt{t^2+1}} D_x-\frac{I x}{\sqrt{t^2+1}} D_y]
\end{align*}

\next
Simply asking for the Petrov type of the spacetime (via the frame $NT$) will classify this metric assuming the parameter $a$ and function $f$ and the spacetime point are all generic.  Generically, the metric is of type II.

\next
{\proc > PetrovType(NT);}

\medskip
\centerline{\it ``II''}
\medskip

Generic choices of $a$ and $f(t)$, however, do not lead to solutions of the Einstein equations. We now investigate the Petrov types which can occur given that $a$ and $f(t)$ are adjusted so that $g$ is an Einstein metric.  It is most efficient to perform all tensor computaions in the anholonomic frame provided by $NT$. To this end we initialize the anholonomic frame defined by $NT$ and define the metric $\eta$ in that frame.

\next
{\proc > FD := FrameData(NT, N): DGsetup(FD):}

\next
{\proc > eta := evalDG(-Theta1 \&s Theta2 + Theta3 \&s Theta4);}

\begin{equation*}
\eta  := -\Theta1 \Theta2 - \Theta2 \Theta1 + \Theta3 \Theta4 + \Theta4 \Theta3
\end{equation*}

We begin by computing the vacuum Einstein equations with cosmological constant. We first define these equations as a tensor field $EFE$, then we extract the components of $EFE$ to create a system of differential equations denoted by $EQ$. The output is suppressed because of its complexity.

\next
{\proc > EFE := evalDG(EinsteinTensor(eta) + Lambda * InverseMetric(eta)):}

\next
{\proc >  EQ := DGinfo(EFE, ``CoefficientSet''):}

\next
The following application of the {\tt PetrovType} command performs the following steps:\footnote{The option {\tt method = ``real''} limits the solving routines to real solutions.}

\begin{enumerate} 
\item From  the given null tetrad, the equations determining the various PetrovTypes are computed. 

\item The Einstein equations are adjoined to this system of equations. 

\item The resulting system of equations is solved for real solutions $a$, $\Lambda$ and $f(t)$,  isolating any exceptional values along the way. 

\end{enumerate}
In this way we find all choices of $a$, $\Lambda$ and $f(t)$ which yield various possible Petrov types under the condition that the metric is an Einstein metric.  The result is presented as a table (PT) whose individual entries are displayed below. 

\next
{\proc > Tetrad := [E1, E2, E3, E4]:}

\next
{\proc > PT := PetrovType(Tetrad, auxiliaryequations = EQ, parameters = [f(t), a, Lambda], method = "real"):}

\next
{\proc > PT[``I''];}
\begin{equation*}
[]
\end{equation*}

\next
{\proc > PT[``II''];}
\begin{equation*}
``generic''
\end{equation*}

\next
{\proc > PT[``III''];}
\begin{equation*}
[a = \half, \Lambda = -\frac{39}{16}, f(t) = \frac{13}{32} t^2+\frac{17}{32}]
\end{equation*}

\next
{\proc > PT[``D''];}
\begin{equation*}
[a = 0, \Lambda = \Lambda, f(t) = (-\frac{1}{6} t^3 \Lambda-\Lambda t-t+\half \frac{\Lambda+2}{t}+C1) \frac{t}{t^2+1}]
\end{equation*}

\next
{\proc > PT[``N''];}
\begin{equation*}
[a = 2, \Lambda = -3, f(t) = \half t^2-\half]
\end{equation*}

\next
{\proc > PT[``O''];}
\begin{equation*}
[[a = 0, \Lambda = -\frac{3}{2}, f(t) = \frac{1}{4}+\frac{1}{4} t^2], [a = -\frac{1}{3}, \Lambda = -\frac{2}{3}, f(t) = \frac{1}{9} t^2-\frac{1}{9}]]
\end{equation*}

The first entry of the table shows that the spacetime defined by  (\ref{1367}) cannot be of Petrov type I.  The second table entry shows (again) that the metric is generically of type II, but is not an Einstein metric in this case.  The remaining entries show that there is a unique Einstein metric of Petrov type III and of type N. There are two Einstein metrics of type O; these are two anti-De Sitter spaces.  Finally, there is a two parameter family of Einstein metrics of Petrov type D. 
That (\ref{1367}) contains Einstein metrics of types III and N is well-documented \cite{Stephani2003}, although our result for the type N case corrects an error in \cite{Stephani2003} for the value of $\Lambda$.    The observation that the family of metrics (\ref{1367}) contains Einstein spaces of type D is apparently new.

\subsection{Adapted tetrads and principal null directions}

Each Petrov type admits a family of adapted null tetrads which put the Newman-Penrose Weyl scalars into a normal form \cite{Penrose1984}.  Such tetrads  can be viewed as optimized for virtually all computations in open regions of spacetime where the Petrov type is constant.  Closely related to these tetrads are the principal null directions, which give important geometrical information about a spacetime.   Here we show \DG\ can be used to find the adapted null tetrads and principal null directions for the G\"odel spacetime.  Computation of adapted null tetrads  and principal null directions both rely upon factorizing the Weyl spinor. The spinor capabilities of \DG\ are detailed in the next section.

We begin by defining the G\"odel metric $g$ and computing its Petrov type.  

\next
{\proc > DGsetup([t, x, y, z], M):}

\next
{\proc >  omega := evalDG(dt  + exp(x)*dz);}

\begin{equation}
\omega := dt + e^{x} dz\nonumber
\end{equation}

\next
{\proc > g := evalDG(-omega \&t omega + dx \&t dx + dy \&t dy  +1/2*exp(2*x)*dz \&t dz );}

\begin{equation}
\label{godel}
g := - dt\ dt + e^{x}  dt\ dz + e^{x}dz\  dt + dx\ dx + dy\ dy + e^{2x} dz\ dz\nonumber
\end{equation}

\next
{\proc > PetrovType(g);}

\begin{equation}
{\it ''D''}
\end{equation}

An obvious orthonormal coframe for the G\"odel metric is given by
\begin{equation}
[\omega, dx, dy, \frac{1}{\sqrt{2}} e^x dz]
\end{equation}
The orthonormal tetrad dual to this coframe and its associated null tetrad can be computed via

\next
{\proc > OT := DualBasis([omega, dx, dy, (1/sqrt(2))*exp(x)*dz]);}
\begin{equation}
OT := [D_{t}, D_{x}, D_{y}, -\sqrt{2} D_{t} + \sqrt{2} e^{-x} D_{z}]
\end{equation}

\next
{\proc > NT := NullTetrad(OT):}

\next
An adapted null tetrad for Petrov type D curvature has its Newman-Penrose Weyl curvature scalars $(\Psi_{0}, \Psi_{1},\dots,\Psi_{4})$ all vanishing save for $\Psi_{2}$. The Weyl curvature scalars can be computed from the frame $NT$ using the \DG\ command {\tt NPCurvatureScalars} which returns a table consisting of the values $(\Psi_{0}, \Psi_{1},\dots,\Psi_{4})$. This computation reveals that the tetrad given above is not adapted to the Petrov type:

\next
{\proc > NPCurvatureScalars(NT, output=[``WeylScalars'']);}

\begin{equation}
table{\rm([``Psi0" = 1/4, ``Psi1" = 0, ``Psi2" = 1/12,  ``Psi3" = 0, ``Psi4" = 1/4 ])}
\end{equation}
To find the optimal tetrads for this Petrov type D spacetime we use the procedure {\proc AdaptedNullTetrad}, applied to $NT$:\footnote{The routine returns two trivially related tetrads; we select the first via {\tt allvalues}.}

\next
{\proc > NTadapted0 := AdaptedNullTetrad(NT, ``D''):}

\next
{\tt >  NTadapted := allvalues(NTadapted0)[1];}

\begin{equation*}
NTadapted := [\sqrt{2}D_t + \sqrt{2} D_y,  \frac{\sqrt{2}}{4} D_t-\frac{\sqrt{2}}{4}D_y, I D_t + \frac{\sqrt{2}}{2} D_x -I e^{-x}D_z,
-I D_t + \frac{\sqrt{2}}{2} D_x + I e^{-x}D_z]
\end{equation*}
We verify that the adapted null tetrad puts the Weyl tensor in normal form.

\next
{\proc > NPCurvatureScalars(NTadapted, output=[``WeylScalars'']);}

\begin{equation}
table{\rm ([``Psi0" = 0, ``Psi1" = 0, ``Psi2" = -1/6,  ``Psi3" = 0, ``Psi4" = 0])}
\end{equation}

For a type D spacetime such as G\"odel the first two legs of the the adapted null tetrad correspond to the principal null directions:

\next
{\proc > allvalues(PrincipalNullDirections(NT, "D"))[1];}

\begin{equation*}
[\sqrt{2}D_t +\sqrt{2} D_y,  \frac{\sqrt{2}}{4} D_t-\frac{\sqrt{2}}{4}D_y]
\end{equation*}

\section{Spinor fields - principal spinors}

\DG\ can be used for the construction and analysis of vector bundles over a manifold.  This capability is used, for example, to provide the jet space functionalities of \DG. Here we briefly show how this capability is used also to perform spinor analysis on a four-dimensional spacetime.  We focus on the problem of factorization of the Weyl spinor of the G\"odel spacetime in terms of principal spinors.

Recall that a spinor field over a four-dimensional spacetime $(M,g)$ (admitting a spin structure) can be viewed as a section of a complex vector bundle over $M$.  This bundle can be viewed as arising by taking tensor products of a complex 2-dimensional vector bundle $E$, with typical fiber $W$, and its complex conjugate bundle $E^\prime$, along with their dual bundles $E^*$ and $E^{\prime*}$. A rank two skew spinor field $\epsilon$ (respectively $\epsilon^\prime$) on $M$ provides at each $x\in M$ an isomorphism between $W$ and $W^*$ (respectively $W^\prime$ and $W^{\prime*}$). The bundle $E\otimes E^\prime$ is equipped with a solder form $\sigma$, which at each $x\in M$ is an isomorphism from $T_xM$ to the Hermitian elements of $W\otimes W^\prime$.  The solder form is the ``square root of the metric'' in the sense that $\sigma^T \circ \sigma = g$, where the transposition operation  is performed with $\epsilon$ and $\epsilon^\prime$.  

To construct spinor fields in \DG\ the first step is to define a chart on $E\otimes E^\prime$, for example

\next
{\proc > DGsetup([t, x, y, z], [ z1, z2, w1, w2]):}

\next
Here coordinates on the base $M$ are $(t,x,y,z)$, (complex) fiber coordinates on $W$ are $(z_1, z_2)$ with complex conjugate coordinates on $W^\prime$ given by $(w_1, w_2)$.  

The epsilon spinors are pre-defined for convenience; one simply has to specify whether one is interested in the covariant spinor fields $\epsilon$ or $\epsilon^\prime$,  or their contravariant versions. For example, the covariant spinors are given by:

\next{\proc > epsilon1 := EpsilonSpinor("spinor", "cov");}

\begin{equation*}
\epsilon1 := dz1 \, dz2 - dz2\, dz1
\end{equation*}

\next{\proc > epsilon2 := EpsilonSpinor("barspinor", "cov");}

\begin{equation*}
\epsilon2 := dw1 \, dw2 - dw2\, dw1
\end{equation*}

A solder form is determined by a choice of orthonormal frame.   Here is an orthonormal frame for the G\"odel metric $g$  given in (\ref{Godel}) and the corresponding solder form, all expressed in the given chart on $E\otimes E^\prime$:

\next
{\proc > OT := evalDG([D\_t, D\_x, D\_y, sqrt(2)*(-D\_t + exp(-x)*D\_z)];}

\begin{equation*}
OT := [D_t, D_x, D_y, -\sqrt{2} D_t + \sqrt{2} e^{-x} D_z]
\end{equation*}

\next
{\proc > sigma := SolderForm(OT);}

\begin{align*}
\sigma := &\frac{\sqrt{2}}{2}dt D_{ z_1} D_{ w_1}+\frac{\sqrt{2}}{2}dt D_{ z_2} D_{ w_2}+\frac{\sqrt{2}}{2}dx D_{ z_1} D_{ w_2}+\frac{\sqrt{2}}{2}dx D_{ z_2} D_{ w_1}
-\frac{I  \sqrt{2} }{2}dy D_{ z_1} D_{ w_2}\\ 
&+\frac{I\sqrt{2} }{2} dy D_{ z_2} D_{ w_1}+\frac{1}{2} e^x (1+ \sqrt{2}) dz D_{ z_1} D_{ w_1}+\frac{1}{2}  e^x ( \sqrt{2}-1) dz D_{ z_2} D_{ w_2}
\end{align*}

\next
The G\"odel metric (\ref{Godel}) can be recovered via contraction of spinor indices in the product of $\sigma$ with itself, a short-cut is provided by the command {\tt SpinorInnerProduct}

\next
{\proc > SpinorInnerProduct(sigma, sigma);}

\begin{equation*}
dt\, dt + e^x dt\, dz - dx\, dx - dy\, dy + e^x dz\, dt + \frac{1}{2} e^{2x} dz\, dz
\end{equation*}

\next
Note that the spinor formalism uses spacetime signature $(+---)$.

The solder form determines a unique {\it spin connection}, $\Gamma$,  which defines a covariant derivative $\nabla$ for an section of any of the (tensor product) spin bundles. This connection is determined from the condition that $\nabla \sigma = 0$, where the Levi-Civita connection compatible with the metric is used to handle parallel transport in the cotangent bundle of $M$.  In \DG\ the connection is defined via

\next
{\proc > Gamma := SpinConnection(sigma):}

\next
Along with the Levi-Civita connection $C$,

\next
{\proc > C := Christoffel(g):}

\next
we can check that the covariant derivative so-defined annihilates the solder form and the epsilon spinors:

\next
{\proc > CovariantDerivative(sigma, C, Gamma);}

\begin{equation}
{0\ dt\, D_{z_1} D_{z_1} dt }
\end{equation}
 
 \next
 {\proc > CovariantDerivative(epsilon1, Gamma);}
 
 \begin{equation}
 0\, dz1\, dz1\, dt
 \end{equation}
 
\next

The spin connection determines a spinorial form of the curvature tensor, which can be computed with the {\tt CurvatureTensor} command. Using the solder form and its inverse this can be converted to a rank-8 spinor field, the {\it curvature spinor}. Using the spinor decomposition of the curvature spinor one can extract the Ricci spinor and the rank-4, totally symmetric {\it Weyl spinor}.  All of these constructions can be performed explicitly, or using pre-defined commands in \DG. For example, the computation of Weyl spinor defined by the solder form $\sigma$ for the G\"odel spacetime is given by

\next
{\proc > W := WeylSpinor(sigma);}

\begin{align*}
W := &-\frac{1}{4} d z_1 d z_1d z_1 d z_1-\frac{1}{12} d z_1 d z_1 d z_2 d z_2-\frac{1}{12} d z_1 d z_2 d z_1 d z_2-\frac{1}{12} d z_1 d z_2 d z_2 d z_1\\
&-\frac{1}{12} d z_1 d z_2 d z_1 d z_2
-\frac{1}{12} d z_1 d z_2 d z_2 d z_1-\frac{1}{12} d z_2 d z_2 d z_1 d z_1-\frac{1}{4}d z_2 d z_2 d z_2 d z_2
\end{align*}

\next

We now consider the problem of factoring the Weyl spinor at a given spacetime point in terms of rank-1 spinors. Any symmetric spinor such as the Weyl spinor can be written as the symmetric product of rank-1 {\it principal spinors} \cite{Penrose1984}
\begin{equation}
\label{factorW}
W  = \alpha\odot\beta\odot\gamma\odot\delta.
\end{equation}
Here we are viewing the Weyl spinor as a covariant spinor, so the principal spinors are covariant rank-1 spinors.  The Petrov type of the spacetime corresponds to the number of distinct spinors featuring in this factorization. The null vectors corresponding to the principal spinors via the solder form are the principal null directions.   \DG\ uses the algorithm described in \cite{Penrose1984} to put the Weyl spinor in normal form, from which it is straightforward to extract the principal spinors.

In our Go\"del example above, the Petrov type is D, so the Penrose normal form is in fact the factorized form.  The principal spinors are not, in general, unique. In the type D case we can demand that the two distinct principal spinors form an oriented, normalized spinor dyad; this reduces the number of possible spinors.  The {\tt FactorWeylSpinor} command produces in the type D case  a list consisting of possible sets of normalized principal spinors in the form $[\alpha, \alpha, \beta, \beta]$  and the overall numerical factor $\eta$ such that the totally symmetric product is of the form:
\begin{equation}
W = \eta\ \alpha\odot  \alpha\odot \beta \odot\beta.
\end{equation}
 We select a single one of these sets to display here:

\next
{\proc > PS, eta := FactorWeylSpinor(W, "D"): }

\next
{\proc > PS:=allvalues(PS[1])[2];}

\begin{equation*}
PS := [I dz_1+dz_2, I dz_1+dz_2, -\frac{1}{2} dz_1-\frac{I}{2}   dz_2, -\frac{1}{2}  dz1-\frac{I}{2} dz2]
\end{equation*}

We can easily check that the spinors thus obtained are in fact principal spinors by taking the symmetric tensor product and comparing with the Weyl spinor $W$:

\next
{\proc > W1 := SymmetrizeIndices(alpha \&t alpha \&t beta \&t beta, [1,2,3,4],  "Symmetric"):}

\next
{\proc > evalDG(W1 - eta * W);}

\begin{equation*}
0\, dz_1 dz_1 dz_1 dz_1
\end{equation*}

\next
We can also verify that the principal spinors thus obtained correspond to the principal null directions obtained in the previous section by computing the null vectors corresponding to the principal spinors via the solder form. This can be done directly via tensor products and contraction, but a shortcut is provided by the {\tt Convert} command, which will convert any spinor with values in $W\otimes W^\prime$ into a spacetime vector.  To illustrate this, the following command take the first principal spinor in $PS$, convert it to a contravariant spinor $PS1$, multiply it by its conjugate spinor and convert the resulting spinor $PNS$ to a (principal null) vector $PNV$:

\next
{\proc > PS1 :=  RaiseLowerSpinorIndices(PS[1], [1]):}

\next
{\proc > PNS := PS1 \&tensor ConjugateSpinor(PS1);}

\begin{equation*}
PNS := D_{z_1}D_{w_1}+ID_{z_1}D_{w_2}-ID_{z_2}D_{w_1}+D_{z_2}D_{w_2}
\end{equation*}

\next
{\proc > PNV := convert(PNS, DGtensor, sigma, [[1,2]]);}

\begin{equation}
\label{PNV}
PNV := \sqrt{2} D_t - \sqrt{2} D_y
\end{equation}
This is indeed one of the principal null directions found in the last section.

\section{Tensorial Invariants}

Besides Killing vector fields there are a number of tensor fields whose existence  can be used to invariantly characterize spacetimes.  Such tensor fields are discussed, {\it e.g.}, in \cite{Stephani2003}. Here we consider a few of these tensorial invariants and show how they can be computed via \DG.

\subsection{Homothetic Vector Fields}

A {\it homothety} of a spacetime $(M,g)$  is a diffeomorphism $\phi\colon M\to M$ and a constant $c\neq0$ such that
\begin{equation}
\phi^{*} g = c g.
\end{equation}
A one parameter family of such diffeomorphisms is generated by a {\it homothetic vector field} $V$, which satisfies
\begin{equation}
\label{Homo}
L_{V} g = (const.) g.
\end{equation}
Evidently, a Killing vector field is a special case of a homothetic vector field, and both are special cases of a conformal Killing vector field.  Modulo Killing vector fields, there is at most one homothetic vector field (up to a scaling). A homothety is a scaling symmetry of the metric; generic spacetime will admit no homotheties.  Indeed, (\ref{Homo}) represents an overdetermined system of PDEs for $V$; such systems are amenable to symbolic analysis.  

As an example, we consider the plane wave spacetimes with  metric given by
\begin{equation}
\label{pw}
g =b(u) \left[ a(u) du \otimes du - P^\prime(u) Q^{\prime}(u) du \odot dv + Q^{\prime}(u)dx \otimes dx + {P^{\prime}(u) }dy \otimes dy\right],
\end{equation}
which is defined in \DG\  via

\next
{\proc > DGsetup([u,v,x,y], M):}

\next
{\proc > g := evalDG(b(u) * (a(u)*du \&t du -   diff(P(u),u)*diff(Q(u),u)*du \&s dv 

+ diff(Q(u), u)*dx \&t dx +  diff(P(u),u)*dy \&t dy):}

\next
Here $a(u)$, $b(u)>0$, $P(u)$ and $Q(u)$ are arbitrary functions such that $P^\prime>0$, $Q^\prime>0$.  Evaluation of the Ricci tensor reveals that this is a vacuum spacetime for an appropriate choice  of $b(u)$, which we shall not need to display.

This metric has a number of remarkable properties which stem from the existence of a continuous homothety \cite{Torre2006}. The corresponding homothetic vector field is obtained in \DG\  via

\next
{\proc > HV := HomothetyVectors(g);}

\begin{align}
HV : =&\Big[ \left(2 v - \int\frac{a(u)}{P^\prime(u) Q^{\prime}(u)}\, du\right) D_{v} + x D_{x} + y D_{y}\Big],\nonumber\\
& \Big[D_{y}, D_{v} + Q(u) D_{y}, D_{x}, x D_{v} + P(u) D_{x}, D_{v}\Big]
\end{align}

\next
The output of {\tt HomothetyVectors} is a pair of lists. The first list contains a non-trivial  homothetic vector field (if any). The second list gives a basis for the vector space of Killing vector fields.   Together, the two lists provide a basis for the infinitesimal generators of the homothety group. For the plane wave metric (\ref{pw}) we see that the homothety group of this spacetime is 6-dimensional and is defined by the choice of functions $P$ and $Q$.  

The corresponding 1-parameter family of homotheties can be computed using the \DG\  {\tt Flow} command.  The output is a list of equations specifying the diffeomorphisms. The one parameter family of homotheties generated by the first vector field in {\proc HVF} is given by:

\next
{\proc > phi := Flow(HV[1][1], s);}

\begin{equation}
\phi := \left[ u = u, \, v = e^{2s}v +  \frac{1}{2} (1- e^{2s})\int\frac{a(u)}{Q^{\prime}(u)}\, du, \,  x = e^{s}x,  \, y = e^{s}y\right]
\end{equation}

\next
This can be verified explicitly by checking  $\phi ^{*}g - e^{2s} g=0$:

\next
{\proc > Pullback(phi, g) \&minus (exp(2*s) \&mult g);}

\begin{equation*}
0 \, du du
\end{equation*}

\next

\subsection{Killing Tensors and Killing-Yano Tensors}

Killing tensors are generalizations of Killing vectors. A Killing tensor on a spacetime $(M,g)$ is a symmetric tensor, $K_{a_{1}\dots a_{p}} = K_{(a_{1}\dots a_{p})}$, satisfying

\begin{equation}
\label{KT}
\nabla_{(a}K_{b_{1}\dots b_{p})} = 0,
\end{equation}
where $\nabla$ is the connection on $M$ compatible with $g$.  Note that a Killing tensor of rank-1 is just the 1-form corresponding via $g$ to a Killing vector field.  The metric $g$ is always a Killing tensor of rank 2. 

A Killing-Yano tensor is a skew tensor, $Y_{a_{1}\dots a_{p}} = Y_{[a_{1}\dots a_{p}]}$, satisfying

\begin{equation}
\nabla_{(a}Y_{b_{1})\dots b_{p}} = 0.
\end{equation}
%
%
We note that if $Y_{ab}$ is a rank-2 Killing-Yano tensor then its square

\begin{equation}
\label{KY2}
K_{ab} = Y_{a}{}^{c} Y_{cb}
\end{equation}
is a Killing tensor.

As with the equations for a Killing vector field or a homothetic vector field, the equations defining Killing tensors and Killing-Yano tensors are overdetermined; generic spacetimes admit no such tensors.   The overdetermined nature of the equations makes them very amenable to analysis using the symbolic PDE solving routines.
As an example, we compute the rank 2 Killing-Yano tensor for the Kerr spacetime. This tensor and its corresponding Killing tensor are  responsible for many important properties of that spacetime \cite{Stephani2003}.  The Killing-Yano tensor for the Kerr spacetime  is obtained as follows.  

We first define the spacetime metric in Boyer-Lindquist coordinates. To keep the expressions manageable, we express the metric in terms of the traditional symbols $\Delta$ and $\rho$, which are then explicitly defined.

\next
{\proc > DGsetup([t, r, theta, phi], M):}

\next
{\proc > g := evalDG(- dt \&t dt + rho\caret2*( 1/Delta * dr \&t dr + dtheta \&t dtheta) + (r\caret2 + a\caret2)*sin(theta)\caret2 *dphi \&t dphi + 2*m*r/rho\caret2 *(a* sin(theta)\caret2 * dphi - dt) \&t (a* sin(theta)\caret2 * dphi - dt));}

\begin{align}
g:=&\frac{2 m r-\rho^2}{\rho^2}dt\ dt-\frac{2 m r a \sin(\theta)^2 }{\rho^2}dt\ d\phi +\frac{\rho^2}{\Delta} dr\ dr+\rho^2 d\theta\ d\theta\nonumber\\
&-\frac{2 m r a \sin(\theta)^2}{\rho^{2}} d\phi \ dt+\frac{\sin(\theta)^2 (2 m r a^2 \sin(\theta)^2+\rho^2 r^2+\rho^2 a^{2}}{\rho^{2}} d\phi\ d\phi
\end{align}

\next
{\proc > definitions := [Delta = r\caret2 - 2*m*r + a\caret2, rho = sqrt(r\caret2 + a\caret2*cos(theta)\caret2)]:}

\begin{equation}
{\it  definitions} := [\Delta = r^{2} - 2mr + a^{2},\ \rho = \sqrt{(r^{2} + a^{2} \cos(\theta)^{2}}]
\end{equation}

\next{\proc > g := eval(g, definitions):}

\next
We then solve for the rank 2 Killing-Yano tensors; the output is a 2-form

\next{\proc > KY := KillingYanoTensors(g, 2);}

\begin{align}
KY := [&-a \cos(\theta) dt \wedge dr+ar\sin(\theta)dt \wedge d\theta-a^2\sin(\theta)^2\cos(\theta)dr \wedge d\phi\nonumber\\
&+r(r^2+a^2)\sin(\theta)d\theta \wedge d\phi]
\end{align}
showing that, up to a multiplicative constant, there is a single Killing-Yano tensor.

The corresponding Killing tensor can now be computed according to (\ref{KY2}) and the Killing tensor equation  (\ref{KT}) can be verified. The expression of the Killing tensor ($KT$) in the coordinate basis is somewhat lengthy; it is computed via the command:

\next
{\proc > KT := ContractIndices(ginv \&tensor KY \&tensor KY, [[4,5]]):}

\next
 A very simple expression of this tensor can be found using a judiciously chosen adapted null tetrad:

\next
{\proc > NT := eval(evalDG([(r\caret2 + a\caret2)/Delta * D\_t + D\_r + a/Delta * D\_phi, (r\caret2 + a\caret2)/2/rho\caret2 * D\_t - Delta/2/rho\caret2 * D\_r + a/2/rho\caret2 * D\_phi, I*a*sin(theta)/sqrt(2)/(r + I*a*cos(theta)) * D\_t + 1/sqrt(2)/(r + I*a*cos(theta)) * D\_theta + I/sin(theta)/sqrt(2)/(r + I*a*cos(theta)) * D\_phi,-I*a*sin(theta)/sqrt(2)/(r - I*a*cos(theta)) * D\_t + 1/sqrt(2)/(r - I*a*cos(theta)) * D\_theta - I/sin(theta)/sqrt(2)/(r - I*a*cos(theta)) * D\_phi]), definitions):}
 
\next
with dual basis computed by

\next
{\proc > DualNT := DualBasis(NT):}

\next
One way to compute the components of $KT$ in this tetrad is to create a basis of symmetric rank-2 tensors via tensor products of the elements of $DualNT$, and then solving for the coefficients of $KT$ expanded in this basis:

\next{\proc > S2 := GenerateSymmetricTensors(DualNT, 2):}

\next{\proc > components := GetComponents(KT, S2);}

\begin{equation*}
components := [0, -2 a^2 cos^2\theta, 0, 0, 0, 0,0, 0, -2 r^2, 0]
\end{equation*}

\next
This indicates that the tensor $KT$ is of the form
\begin{equation}
KT =  -2 a^2 cos^2\theta k \odot l - 2 r^2 m\odot \overline m,
\end{equation}
where
\begin{equation*}
(k,l,m,\overline m) = (DualNT[1], DualNT[2], DualNT[3], DualNT[4])
\end{equation*}

To verify that $KT$ is in fact a Killing tensor, we simply check that the symmetrized covariant derivative vanishes, as follows.

\next
{\proc > ginv := InverseMetric(g): C := Christoffel(g):}

\next
{\proc > DKT := CovariantDerivative(KT, C):}

\next
{\proc > SymmetrizeIndices(DKT, [1,2,3], ``Symmetric'');}

\begin{equation*}
0\ dt\ dt\ dt 
\end{equation*}

An important question that arises when computing Killing tensors is whether they are ``reducible'' in the sense of being constructed from (linear combinations over {\bf R} of) the symmetric products of lower rank Killing tensors.  It is possible to construct various symbolic algorithms which answer this question. Within \DG, a good strategy is to exploit the command {\tt GetComponents}, just used above, which gives the components of any element of a vector space in terms of a given set of elements of that space. To illustrate this, we consider the rank 2 Killing tensors of the G\"odel spacetime. With the spacetime as defined in (\ref{godel}) we compute the rank-2 Killing tensors via

\next{\proc > KT := KillingTensors(g,2):}

\next
We have suppressed the lengthy output, which is a list of rank-2 tensor fields forming a basis for the vector space $K$ (over {\bf R}) of solutions to (\ref{KT}). The dimension of $KT$ is obtained by asking how many elements are in this list:

\next{\proc > nops(K);}

\begin{equation*}
15
\end{equation*}

\next
To compute the ``trivial'' Killing tensors, we compute a basis $KV$ of rank-1 Killing tensors (corresponding to the 5 Killing vector fields of the G\"odel metric) and a basis $B$ for the algebra $S$ of rank-2 symmetric tensors generated by them. Again we suppress the lengthy output.

\next
{\proc > KV := KillingTensors(g,1):}

\next{\proc > B := SymmetricProductsOfKillingTensors(KV, 2):}

\next
We then check to see which elements of $KT$ are in the algebra $S$, by checking if they are in the span of $B$ with real coefficients. Rather than displaying the components of $KT$ in the basis $B$, we simply ask if the Killing tensors are in the span of $B$:

\next 
{\proc > GetComponents(KT, B, method=``real'', trueorfalse=``on'');}

\begin{equation}
true
\end{equation}

The result ``true'' indicates that {\it all} the rank-2 Killing tensors are spanned by the rank-1 Killing tensors, in agreement with results of Cook \cite{Cook2009}.  This, of course, could have been deduced from the fact that the dimension of $KT$ is 15, which is also the dimension of the symmetric algebra of rank-2 tensors generated by the 5 Killing vector fields.   A similar (but more CPU intensive) analysis establishes that there are no non-trivial rank-3 Killing tensors for the G\"odel spacetime.

%
%

%

\section{Symmetry Reduction of Field Equations}

Probably the most effective general method for solving non-linear field equations is to look for solutions admitting a prescribed group of symmetries. Given a system of differential equations for fields on a manifold $M$ which are invariant under a transformation group $G$, one can convert the problem of finding $G$-invariant solutions to that of solving a  generally simpler set of reduced equations for fields on $M/G$. The \DG\  package provides an environment in which structures from differential geometry, Lie algebras and groups, and fiber bundles are tightly integrated.  As such it is well-suited for geometric analysis of field equations and, in particular, the use of symmetry reduction to analyze group-invariant solutions of field equations \cite{AFT2000}.  Here we focus on two important tasks which arise when analyzing a symmetry reduction of a field theory: (i) characterizing the most general group invariant field, (ii) calculation of the induced symmetry group of the reduced field equations.  We illustrate this with the symbolic analysis of the problem of finding gravitational plane waves, which can be defined as vacuum spacetimes admitting a particular class of 5 dimensional isometry groups (see \cite{Torre2006}, and references therein).

The manifold is denoted by $M$ with coordinates $(u,v,x,y)$. The desired isometry group action is the  5-dimensional group $G$ of diffeomorphisms  of $M$ generated by the vector fields
\begin{equation}
\label{PWSA}
X_{1} = \frac{\partial}{\partial v}, \ X_{2} = \frac{\partial}{\partial x}, \ X_{3} = \frac{\partial}{\partial y},\ 
X_{4}  = x \frac{\partial}{\partial v} + P(u) \frac{\partial}{\partial x},\ X_{5}  = y \frac{\partial}{\partial v} + Q(u) \frac{\partial}{\partial y},
\end{equation}
where $P(u)$ and $Q(u)$ are any functions such that $P^\prime(u)>0$, $Q^\prime(u)>0$.  Every gravitational plane wave admits $G$ as an isometry group for some choice of the functions $P$ and $Q$; these functions determine the amplitude and polarization of the wave.

We define the spacetime manifolds and the infinitesimal generators of $G$ as follows.

\next
{\proc > DGsetup([u,v,x,y], M):}

\next
{\proc > 
X1 := D\_v: X2 := D\_x: X3 := D\_y:             X4 := evalDG(x*D\_v + P(u)*D\_x):

X5 := evalDG(y*D\_v + Q(u)*D\_y):}

\next
{\proc > KV := [X1, X2, X3, X4, X5]:}

\next

Our first task is to characterize the $G$-invariant spacetime metrics, that is, to find the general form of rank-2 symmetric tensor fields which are invariant under the group action generated by the vector fields $KV$. This can be done using the command {\tt InvariantGeometricObjectFields}, which solves the linear system of PDEs defining any group invariant field.   For example, here we find the invariant functions:\footnote{The second argument of this command is a list which defines a basis (with respect to linear combinations with coefficients given by functions on $M$) for the invariant tensor fields of the desired type.}

\next
{\proc > InvariantGeometricObjectFields(KV, [1]);}

\begin{equation*}
F1(u)
\end{equation*}
Thus all invariant functions are functions of the invariant $u$. In particular, the group orbits are the hypersurfaces $u=const.$ and the reduced spacetime manifold is $M/G \approx {\bf R}$ with coordinate $u$.

Similarly we can compute the most general symmetric tensor of type $\left(0\atop2\right)$ which is invariant under $G$.  This determines the ``reduced fields'' from which group invariant solutions of field equations are obtained. First we compute a basis $S2$ for symmetric tensors of type $\left(0\atop2\right)$, then we compute the $G$ invariant fields :

\next
{\proc > S2:=GenerateSymmetricTensors([du, dv, dx, dy], 2):}

\next
{\proc > S2invariant := InvariantGeometricObjectFields(KV, S2);}

\begin{align}
S2invariant := &F1(u) \Big(-\frac{d}{du}P(u) \frac{d}{du} Q(u) (du\, dv + dv\, du)
+ \frac{d}{du}Q(u) dx\, dx + \frac{d}{du}P(u) dy\, dy\Big)\nonumber\\ 
&+ F2(u) du\, du
\end{align}



\next
Thus the set of $G$-invariant symmetric tensors of type $\left(0\atop2\right)$ is spanned by two quadratic forms:
\begin{equation}
\label{Q}
{\cal Q}_{1} = du \otimes du, \quad {\cal Q}_{2} = -Q^{\prime}(u) P^{\prime}(u) du \odot dv + Q^{\prime}(u) dx \otimes dx + P^{\prime}(u) dy \otimes dy.
\end{equation}
 The most general $G$-invariant metric takes the form
\begin{equation}
\label{ginv}
g = a(u) {\cal Q}_{1} + b(u) {\cal Q}_{2},
\end{equation}
where $b(u)>0$.  The symmetry reduced fields on $M/G \approx {\bf R}$ are $a(u)$ and $b(u)$.  


One obtains symmetry reduced equations for $a(u)$ and $b(u)$ by substituting (\ref{ginv}) into the relevant field equations, {\it e.g.}, the Einstein equations.  Generally covariant field equations such as the Einstein equations have the group \DiffM\ of diffeomorphisms of $M$ (acting by pull-back on the metric) as a symmetry group.  The full group of diffeomorphisms does not act on the space of metrics (\ref{ginv}) so \DiffM\ cannot be a symmetry group of the symmetry reduced equations. The subgroup of \DiffM\ which {\it does} act on the $G$-invariant metrics and which will also be a symmetry group of the reduced equations is the normalizer of $G$ within \DiffM\ \cite{AFT2000}.  Thus the normalizer is the {\it residual symmetry group} after symmetry reduction. The normalizer of $G$ in \DiffM\ is the smallest subgroup of \DiffM\ containing $G$ as a normal subgroup.  Working infinitesimally, the Lie algebra of the normalizer can be computed as the smallest algebra of vector fields containing the isometry algebra generated by $KV$ as an ideal. These vector fields are obtained by solving a large system of linear PDEs which are amenable to symbolic analysis. 

The \DG command {\tt InfinitesimalPseudoGroupNormalizer} solves these PDEs and thus computes the Lie algebra of the normalizer within the Lie algebra of all vector fields.\footnote{The default for this computation is to compute the normalizer within the set $\Gamma$ of all vector fields. Other choices for $\Gamma$ can be made.} In the present example we get 

\next
{\proc > Nor := InfinitesimalPseudoGroupNormalizer(KV);}

\begin{equation*}
Nor := (2 C1\, v + F1(u)) D_{v} + C1\, x\, D_{x} + C1\, y\, D_{y}
\end{equation*}

\next
Thus the residual symmetry group consists of a scaling of $(v, x, y)$ and a $u$ dependent translation of $v$.  These transformations can be computed explicitly using the \DG\ {\tt Flow} command as follows.  First, for simplicity, we break the vector field {\it Nor} into its translational and scaling parts, respectively:

\next{\proc > NorTrans := eval(Nor, {C1=0, F1(u) = f(u)});}
\begin{equation*}
NorTrans := f(u) D_{v}
\end{equation*}

\next
{\proc > NorScale := eval(Nor, {C1=1, F1(u) = 0});}
\begin{equation*}
NorScale := 2v\, D_{v} + x\, D_{x} + y \, 
D_{y}
\end{equation*}

\next
Then we compute each of  the flows using parameters $\alpha$ and $\beta$ respectively. The result is two sets of transformations, specified as relations between the original and transformed coordinate values:

\next
{\proc > Trans := Flow(NorTrans,  alpha);}

\begin{equation*}
Trans := [u=u, v= v + \alpha f(u), x=x, y=y]
\end{equation*}

\next
{\proc > Scale := Flow(NorScale,  beta);}

\begin{equation*}
Trans := [u=u, v= e^{2\beta}v, x=e^{\beta}x, y=e^{\beta}y]
\end{equation*}

\next
Any generally covariant field theory (not just general relativity, or a theory of a metric) will admit these transformations as symmetries upon symmetry reduction by the symmetry group defined by (\ref{PWSA}). 

We can verify that the residual symmetry group does indeed act on the $G$-invariant metric (\ref{ginv}) via the {\tt Pullback} command. Defining {\tt Q1} and {\tt Q2} as in (\ref{Q}) we have

\next
{\proc > Pullback(Trans, Q1);}

\begin{equation*}
du\, du
\end{equation*}

\next
{\proc > Pullback(Trans, Q2)}

\begin{equation*}
-2 \alpha {d P(u)\over du} {d Q(u)\over u} {d f(u)\over u} du\, du
- {d P(u)\over du} {d Q(u)\over du} (du\, dv + dv\, du) + {d Q(u)\over du} dx\, dx + {d P(u)\over du} dy\, dy
\end{equation*}

\next
{\proc > Pullback(Scale, Q1);}

\begin{equation*}
du\, du
\end{equation*}

\next
{\proc > Pullback(Scale, Q2);}

\begin{equation*}
e^{2\beta}(- {d P(u)\over du} {d Q(u)\over du} (du\, dv + dv\, du) + {d Q(u)\over du} dx\, dx + {d P(u)\over du} dy\, dy)
\end{equation*}

\next
Thus the action of the residual group on the functions $(a(u), b(u))$ defining the $G$-invariant metrics can be expressed as
\begin{equation}
a(u) \longrightarrow a(u) - 2\alpha P^{\prime}(u) Q^{\prime}(u) f^{\prime}(u) b(u),\quad\quad
b(u) \longrightarrow e^{2\beta} b(u).
\end{equation}

\next
From this action of the residual symmetry group we see that the function $a(u)$ can be ``gauged'' to zero via {\it Trans}. We can also see that, with an appropriate choice of $f(u)$, the residual group includes a homothety of the metric (\ref{ginv}), as discussed in \S6.1.

\appendix

\section{DifferentialGeometry Commands}
 
The following is a list of all commands in \DG\ and its various sub-packages. 

\subsection{DifferentialGeometry package}

\begin{maplelatex}\begin{Maple Bullet Item}{
\&minus:  find the difference between two vectors, differential forms or tensors.}\end{Maple Bullet Item}
\end{maplelatex}
\begin{maplelatex}
\begin{Maple Bullet Item}{
\&mult:  multiply a vector, differential form or tensor by a Maple expression.}\end{Maple Bullet Item}
\end{maplelatex}
\begin{maplelatex}\begin{Maple Bullet Item}{
\&plus:  add two vectors, differential forms or tensors.}\end{Maple Bullet Item}
\end{maplelatex}
\begin{maplelatex}\begin{Maple Bullet Item}{
\&tensor:  calculate the tensor product of two tensors.}\end{Maple Bullet Item}
\end{maplelatex}
\begin{maplelatex}\begin{Maple Bullet Item}{
\&wedge:  calculate the exterior product of two differential forms.}\end{Maple Bullet Item}
\end{maplelatex}
\begin{maplelatex}\begin{Maple Bullet Item}{
Annihilator: find the subspace of vectors (or 1-forms) whose interior product with a given list of 1-forms (vectors) vanish.}\end{Maple Bullet Item}
\end{maplelatex}
\begin{maplelatex}\begin{Maple Bullet Item}{
ApplyTransformation: evaluate a transformation at a point.}\end{Maple Bullet Item}
\end{maplelatex}
\begin{maplelatex}\begin{Maple Bullet Item}{
ChangeFrame: change the current or active frame.}\end{Maple Bullet Item}
\end{maplelatex}
\begin{maplelatex}\begin{Maple Bullet Item}{
ComplementaryBasis: extend a basis for subspace to a basis for larger subspace.}\end{Maple Bullet Item}
\end{maplelatex}
\begin{maplelatex}\begin{Maple Bullet Item}{
ComposeTransformations: compose a sequence of two or more transformations.}\end{Maple Bullet Item}
\end{maplelatex}
\begin{maplelatex}\begin{Maple Bullet Item}{
Convert: change the presentations of various geometric objects.}\end{Maple Bullet Item}
\end{maplelatex}
\begin{maplelatex}\begin{Maple Bullet Item}{
DeRhamHomotopy:  the homotopy operator for the exterior derivative operator (the de Rham complex).}\end{Maple Bullet Item}
\end{maplelatex}
\begin{maplelatex}\begin{Maple Bullet Item}{
DGbasis: select a maximal linearly independent list of elements from a list of vectors, forms or tensors.}\end{Maple Bullet Item}
\end{maplelatex}
\begin{maplelatex}\begin{Maple Bullet Item}{
DGsetup:  initialize a coordinate system, frame, or Lie algebra.}\end{Maple Bullet Item}
\end{maplelatex}
\begin{maplelatex}\begin{Maple Bullet Item}{
DGzip: form a linear combination, wedge product or tensor product of a list of vectors, forms or tensors.}\end{Maple Bullet Item}
\end{maplelatex}
\begin{maplelatex}\begin{Maple Bullet Item}{
DualBasis: calculate the dual basis to a given basis of vectors or 1-forms.}\end{Maple Bullet Item}
\end{maplelatex}
\begin{maplelatex}\begin{Maple Bullet Item}{
evalDG: evaluate a  \textbf{DifferentialGeometry} expression.}\end{Maple Bullet Item}
\end{maplelatex}
\begin{maplelatex}\begin{Maple Bullet Item}{
ExteriorDerivative: take the exterior derivative of a differential form.}\end{Maple Bullet Item}
\end{maplelatex}
\begin{maplelatex}\begin{Maple Bullet Item}{
Flow: calculate the 1-parameter group of diffeomorphisms (the flow) of a vector field.}\end{Maple Bullet Item}
\end{maplelatex}
\begin{maplelatex}\begin{Maple Bullet Item}{
FrameData: calculate the structure equations for a generic (anholomonic) frame.}\end{Maple Bullet Item}
\end{maplelatex}
\begin{maplelatex}\begin{Maple Bullet Item}{
GroupActions:  a package for Lie groups and group actions on manifolds.}\end{Maple Bullet Item}
\end{maplelatex}
\begin{maplelatex}\begin{Maple Bullet Item}{
Hook: the interior product of a vector or a list of vectors with a differential form.}\end{Maple Bullet Item}
\end{maplelatex}
\begin{maplelatex}\begin{Maple Bullet Item}{
InfinitesimalTransformation: compute the Lie algebra of infinitesimal generators for an action of a Lie group on a manifold.}\end{Maple Bullet Item}
\end{maplelatex}
\begin{maplelatex}\begin{Maple Bullet Item}{
IntegrateForm:  evaluate a \textbf{p}-fold iterated integral of a differential \textbf{p}-form.}\end{Maple Bullet Item}
\end{maplelatex}
\begin{maplelatex}\begin{Maple Bullet Item}{
IntersectSubspaces:  find the intersection of a list of vector subspaces of vectors, forms or tensors.}\end{Maple Bullet Item}
\end{maplelatex}
\begin{maplelatex}\begin{Maple Bullet Item}{
InverseTransformation: find the inverse of a transformation.}\end{Maple Bullet Item}
\end{maplelatex}
\begin{maplelatex}\begin{Maple Bullet Item}{
JetCalculus:  a package for the variational calculus on jet spaces.}\end{Maple Bullet Item}
\end{maplelatex}
\begin{maplelatex}\begin{Maple Bullet Item}{
Library:  a package of databases of Lie algebras, vector field systems, and differential equations.}\end{Maple Bullet Item}
\end{maplelatex}
\begin{maplelatex}\begin{Maple Bullet Item}{
LieAlgebras:  a package for the symbolic analysis of Lie algebras.}\end{Maple Bullet Item}
\end{maplelatex}
\begin{maplelatex}\begin{Maple Bullet Item}{
LieBracket: calculate the Lie bracket of two vector fields.}\end{Maple Bullet Item}
\end{maplelatex}
\begin{maplelatex}\begin{Maple Bullet Item}{
LieDerivative:  calculate the Lie derivative of a vector field, differential form or tensor with respect to a vector field.}\end{Maple Bullet Item}
\end{maplelatex}
\begin{maplelatex}\begin{Maple Bullet Item}{
GetComponents: find the coefficients of a vector, differential form or tensor with respect to a list of vectors, differential forms or tensors.}\end{Maple Bullet Item}
\end{maplelatex}
\begin{maplelatex}\begin{Maple Bullet Item}{
Preferences: set worksheet preferences for the  \textbf{DifferentialGeometry} package.}\end{Maple Bullet Item}
\end{maplelatex}
\begin{maplelatex}\begin{Maple Bullet Item}{
Pullback: pullback a differential \textbf{p}-form by the Jacobian of a transformation.}\end{Maple Bullet Item}
\end{maplelatex}
\begin{maplelatex}\begin{Maple Bullet Item}{
PullbackVector: find (if possible) a vector field whose pushforward by the Jacobian of a given transformation is a given vector field.}\end{Maple Bullet Item}
\end{maplelatex}
\begin{maplelatex}\begin{Maple Bullet Item}{
Pushforward: pushforward a vector or a vector field by the Jacobian of a transformation.}\end{Maple Bullet Item}
\end{maplelatex}
\begin{maplelatex}\begin{Maple Bullet Item}{
RemoveFrame: remove a frame from a Maple session.}\end{Maple Bullet Item}
\end{maplelatex}
\begin{maplelatex}\begin{Maple Bullet Item}{
Tensor:  a package for tensor analysis within the  \textbf{DifferentialGeometry} environment.}\end{Maple Bullet Item}
\end{maplelatex}
\begin{maplelatex}\begin{Maple Bullet Item}{
Tools:  a small utility package for  \textbf{DifferentialGeometry} .}\end{Maple Bullet Item}
\end{maplelatex}
\begin{maplelatex}\begin{Maple Bullet Item}{
Transformation: create a transformation or mapping from one manifold to another.}\end{Maple Bullet Item}
\end{maplelatex}

\subsection{GroupActions sub-package}

\begin{maplelatex}\end{maplelatex}
\begin{maplelatex}\begin{Maple Bullet Item}{
Action: find the action of a solvable Lie group on a manifold from its infinitesimal generators.}\end{Maple Bullet Item}
\end{maplelatex}
\begin{maplelatex}\begin{Maple Bullet Item}{
InfinitesimalSymmetriesOfGeometricObjectFields: find the infinitesimal symmetries (vector fields) for a collection of vector fields, differential forms or tensors.}\end{Maple Bullet Item}
\end{maplelatex}
\begin{maplelatex}\begin{Maple Bullet Item}{
InvariantGeometricObjectFields: find the vector fields, differential forms, or tensors which are invariant with respect to a Lie algebra of vector fields.}\end{Maple Bullet Item}
\end{maplelatex}
\begin{maplelatex}\begin{Maple Bullet Item}{
InvariantVectorsAndForms: calculate a basis of left and right invariant vector fields and differential 1-forms on a Lie group.}\end{Maple Bullet Item}
\end{maplelatex}
\begin{maplelatex}\begin{Maple Bullet Item}{
IsotropyFilteration: find the infinitesimal isotropy filtration for a Lie algebra of vector fields.}\end{Maple Bullet Item}
\end{maplelatex}
\begin{maplelatex}\begin{Maple Bullet Item}{
IsotropySubalgebra: find the infinitesimal isotropy subalgebra of a Lie algebra of vector fields and infinitesimal isotropy representation.}\end{Maple Bullet Item}
\end{maplelatex}
\begin{maplelatex}\begin{Maple Bullet Item}{
LieGroup: create a module defining a Lie group.}\end{Maple Bullet Item}
\end{maplelatex}
\begin{maplelatex}\begin{Maple Bullet Item}{
LiesThirdTheorem: find a Lie algebra of pointwise independent vector fields with prescribed structure equations (solvable algebras only).}\end{Maple Bullet Item}
\end{maplelatex}
\begin{maplelatex}\begin{Maple Bullet Item}{
MovingFrames:  a small package for the method of moving frames.}\end{Maple Bullet Item}
\end{maplelatex}

\subsection{JetCalculus sub-package}

\begin{maplelatex}\begin{Maple Bullet Item}{
AssignTransformationType: assign a type (projectable, point, contact, ...) to a transformation.}\end{Maple Bullet Item}
\end{maplelatex}
\begin{maplelatex}\begin{Maple Bullet Item}{
AssignVectorType: assign a type (projectable, point, contact, ...) to a vector.}\end{Maple Bullet Item}
\end{maplelatex}
\begin{maplelatex}\begin{Maple Bullet Item}{
DifferentialEquationData: create a data structure for a system of differential equations.}\end{Maple Bullet Item}
\end{maplelatex}
\begin{maplelatex}\begin{Maple Bullet Item}{
EulerLagrange: calculate the Euler-Lagrange equations for a Lagrangian.}\end{Maple Bullet Item}
\end{maplelatex}
\begin{maplelatex}\begin{Maple Bullet Item}{
EvolutionaryVector: find the evolutionary part of a vector field.}\end{Maple Bullet Item}
\end{maplelatex}
\begin{maplelatex}\begin{Maple Bullet Item}{
GeneralizedLieBracket: find the Lie bracket of two generalized vector fields.}\end{Maple Bullet Item}
\end{maplelatex}
\begin{maplelatex}\begin{Maple Bullet Item}{
GeneratingFunctionToContactVector: find the contact vector field defined by a generating function.}\end{Maple Bullet Item}
\end{maplelatex}
\begin{maplelatex}\begin{Maple Bullet Item}{
HigherEulerOperators: apply the higher Euler operators to a function or a differential bi-form.}\end{Maple Bullet Item}
\end{maplelatex}
\begin{maplelatex}\begin{Maple Bullet Item}{
HorizontalExteriorDerivative:  calculate the horizontal exterior derivative of a bi-form on a jet space.}\end{Maple Bullet Item}
\end{maplelatex}
\begin{maplelatex}\begin{Maple Bullet Item}{
HorizontalHomotopy: apply the horizontal homotopy operator to a bi-form on a jet space.}\end{Maple Bullet Item}
\end{maplelatex}
\begin{maplelatex}\begin{Maple Bullet Item}{
IntegrationByParts: apply the integration by parts operator to a differential bi-form.}\end{Maple Bullet Item}
\end{maplelatex}
\begin{maplelatex}\begin{Maple Bullet Item}{
Noether:  find the conservation law for the  Euler-Lagrange equations from a given symmetry of the Lagrangian}\end{Maple Bullet Item}
\end{maplelatex}
\begin{maplelatex}\begin{Maple Bullet Item}{
ProjectedPullback: pullback a differential bi-form of type \textbf{(r, s)}by a transformation to a differential bi-form of type \textbf{(r, s)}.}\end{Maple Bullet Item}
\end{maplelatex}
\begin{maplelatex}\begin{Maple Bullet Item}{
ProjectionTransformation: construct the canonical projection map between jet spaces of a fiber bundle.}\end{Maple Bullet Item}
\end{maplelatex}
\begin{maplelatex}\begin{Maple Bullet Item}{
Prolong: prolong a jet space, vector field, transformation, or differential equation to a higher order jet space.}\end{Maple Bullet Item}
\end{maplelatex}
\begin{maplelatex}\begin{Maple Bullet Item}{
PushforwardTotalVector: pushforward a total vector field by a transformation.}\end{Maple Bullet Item}
\end{maplelatex}
\begin{maplelatex}\begin{Maple Bullet Item}{
TotalDiff: take the total derivative of an expression, a differential form or a contact form.}\end{Maple Bullet Item}
\end{maplelatex}
\begin{maplelatex}\begin{Maple Bullet Item}{
TotalJacobian: find the Jacobian of a transformation using total derivatives.}\end{Maple Bullet Item}
\end{maplelatex}
\begin{maplelatex}\begin{Maple Bullet Item}{
TotalVector: find the total part of a vector field.}\end{Maple Bullet Item}
\end{maplelatex}
\begin{maplelatex}\begin{Maple Bullet Item}{
VerticalExteriorDerivative: calculate the vertical exterior derivative of a bi-form on a jet space.}\end{Maple Bullet Item}
\end{maplelatex}
\begin{maplelatex}\begin{Maple Bullet Item}{
VerticalHomotopy: apply the vertical homotopy operator to a bi-form on a jet space.}\end{Maple Bullet Item}
\end{maplelatex}
\begin{maplelatex}\begin{Maple Bullet Item}{
ZigZag:  lift a \textbf{dH}closed form on a jet space to a \textbf{d}closed form.}\end{Maple Bullet Item}
\end{maplelatex}

\subsection{LieAlgebras sub-package}

\begin{maplelatex}\begin{Maple Bullet Item}{
Adjoint: find the Adjoint Matrix for a vector in a Lie algebra.}\end{Maple Bullet Item}
\end{maplelatex}
\begin{maplelatex}\begin{Maple Bullet Item}{
AdjointExp: find the Exponential of the Adjoint Matrix for a vector in a Lie algebra.}\end{Maple Bullet Item}
\end{maplelatex}
\begin{maplelatex}\begin{Maple Bullet Item}{
ApplyHomomorphism: apply a Lie algebra homomorphism to a vector, form or tensor.}\end{Maple Bullet Item}
\end{maplelatex}
\begin{maplelatex}\begin{Maple Bullet Item}{
BracketOfSubspaces:  find the subspace generated by the bracketing of two subspaces.}\end{Maple Bullet Item}
\end{maplelatex}
\begin{maplelatex}\begin{Maple Bullet Item}{
Center: find the center of a Lie algebra.}\end{Maple Bullet Item}
\end{maplelatex}
\begin{maplelatex}\begin{Maple Bullet Item}{
Centralizer: find the centralizer of a list of vectors.}\end{Maple Bullet Item}
\end{maplelatex}
\begin{maplelatex}\begin{Maple Bullet Item}{
ChangeLieAlgebraTo: change the current frame to the frame for a Lie algebra.}\end{Maple Bullet Item}
\end{maplelatex}
\begin{maplelatex}\begin{Maple Bullet Item}{
Complexify: find the complexification of a Lie algebra.}\end{Maple Bullet Item}
\end{maplelatex}
\begin{maplelatex}\begin{Maple Bullet Item}{
Decompose: decompose a Lie algebra into a direct sum of indecomposable Lie algebras.}\end{Maple Bullet Item}
\end{maplelatex}
\begin{maplelatex}\begin{Maple Bullet Item}{
Derivations: find the inner and/or Outer derivations of a Lie algebra.}\end{Maple Bullet Item}
\end{maplelatex}
\begin{maplelatex}\begin{Maple Bullet Item}{
DerivedAlgebra: find the derived algebra of a Lie algebra.}\end{Maple Bullet Item}
\end{maplelatex}
\begin{maplelatex}\begin{Maple Bullet Item}{
DirectSum: create the direct sum of a list of Lie algebras.}\end{Maple Bullet Item}
\end{maplelatex}
\begin{maplelatex}\begin{Maple Bullet Item}{
Extension: calculate a right or a central extension of a Lie algebra.}\end{Maple Bullet Item}
\end{maplelatex}
\begin{maplelatex}\begin{Maple Bullet Item}{
GeneralizedCenter:  calculate the generalized center of an ideal in a Lie algebra.}\end{Maple Bullet Item}
\end{maplelatex}
\begin{maplelatex}\begin{Maple Bullet Item}{
HomomorphismSubalgebras: find the kernel or image of a Lie algebra homomorphism.}\end{Maple Bullet Item}
\end{maplelatex}
\begin{maplelatex}\begin{Maple Bullet Item}{
JacobsonRadical: find the Jacobson radical for a matrix Lie algebra.}\end{Maple Bullet Item}
\end{maplelatex}
\begin{maplelatex}\begin{Maple Bullet Item}{
Killing: find the Killing form of a Lie algebra.}\end{Maple Bullet Item}
\end{maplelatex}
\begin{maplelatex}\begin{Maple Bullet Item}{
LeviDecomposition: compute the Levi decomposition of a Lie algebra.}\end{Maple Bullet Item}
\end{maplelatex}
\begin{maplelatex}\begin{Maple Bullet Item}{
LieAlgebraCohomology: a package for computing Lie algebra cohomology.}\end{Maple Bullet Item}
\end{maplelatex}
\begin{maplelatex}\begin{Maple Bullet Item}{
LieAlgebraData: convert different realizations of a Lie algebra to a Lie algebra data structure.}\end{Maple Bullet Item}
\end{maplelatex}
\begin{maplelatex}\begin{Maple Bullet Item}{
LieAlgebraRepresentations:  a package for the symbolic analysis of representations of Lie algebras.}\end{Maple Bullet Item}
\end{maplelatex}
\begin{maplelatex}\begin{Maple Bullet Item}{
MatrixAlgebras: create a Lie algebra data structure for a matrix Lie algebra.}\end{Maple Bullet Item}
\end{maplelatex}
\begin{maplelatex}\begin{Maple Bullet Item}{
MatrixCentralizer: find the matrix centralizer of a list of matrices.}\end{Maple Bullet Item}
\end{maplelatex}
\begin{maplelatex}\begin{Maple Bullet Item}{
MinimalIdeal: find the smallest ideal containing a given set of vectors.}\end{Maple Bullet Item}
\end{maplelatex}
\begin{maplelatex}\begin{Maple Bullet Item}{
MinimalSubalgebra: find the smallest ideal containing a given set of vectors.}\end{Maple Bullet Item}
\end{maplelatex}
\begin{maplelatex}\begin{Maple Bullet Item}{
MultiplicationTable: display the multiplication table of a Lie algebra.}\end{Maple Bullet Item}
\end{maplelatex}
\begin{maplelatex}\begin{Maple Bullet Item}{
Nilradical: find the nilradical of a Lie algebra.}\end{Maple Bullet Item}
\end{maplelatex}
\begin{maplelatex}\begin{Maple Bullet Item}{
Query: check various properties of a Lie algebra, subalgebra, or transformation.}\end{Maple Bullet Item}
\end{maplelatex}
\begin{maplelatex}\begin{Maple Bullet Item}{
QuotientAlgebra: create the Lie algebra data structure for a quotient algebra of a Lie algebra by an ideal.}\end{Maple Bullet Item}
\end{maplelatex}
\begin{maplelatex}\begin{Maple Bullet Item}{
Radical: find the radical of a Lie algebra.}\end{Maple Bullet Item}
\end{maplelatex}
\begin{maplelatex}\begin{Maple Bullet Item}{
SemiDirectSum: create the semi-direct product of two Lie algebras.}\end{Maple Bullet Item}
\end{maplelatex}
\begin{maplelatex}\begin{Maple Bullet Item}{
Series: find the derived series, lower central series, upper central series of a Lie algebra or a Lie subalgebra.}\end{Maple Bullet Item}
\end{maplelatex}
\begin{maplelatex}\begin{Maple Bullet Item}{
SubalgebraNormalizer: find the normalizer of a subalgebra.}\end{Maple Bullet Item}
\end{maplelatex}

\subsection{Tensor sub-package}

{ \textbf{Commands for the algebraic manipulation of tensors}}

\begin{maplelatex}\begin{Maple Bullet Item}{
CanonicalTensors: create various standard tensors.}\end{Maple Bullet Item}
\end{maplelatex}
\begin{maplelatex}\begin{Maple Bullet Item}{
ContractIndices: contract the indices of a tensor.}\end{Maple Bullet Item}
\end{maplelatex}
\begin{maplelatex}\begin{Maple Bullet Item}{
Convert/DGspinor:  convert a tensor to a spinor.}\end{Maple Bullet Item}
\end{maplelatex}
\begin{maplelatex}\begin{Maple Bullet Item}{
Convert/DGtensor:  convert an array, vector, p-form, spinor ... to a  tensor.}\end{Maple Bullet Item}
\end{maplelatex}
\begin{maplelatex}\begin{Maple Bullet Item}{
DGGramSchmidt:  construction an orthonormal basis of vector, forms, tensors with respect to a metric.}\end{Maple Bullet Item}
\end{maplelatex}
\begin{maplelatex}\begin{Maple Bullet Item}{
GenerateSymmetricTensors: generate a list of symmetric tensors from a list of tensors.}\end{Maple Bullet Item}
\end{maplelatex}
\begin{maplelatex}\begin{Maple Bullet Item}{
GenerateTensors: generate a list of tensors from a list of lists of tensors.}\end{Maple Bullet Item}
\end{maplelatex}
\begin{maplelatex}\begin{Maple Bullet Item}{
HodgeStar: apply the Hodge star operator to a differential form.}\end{Maple Bullet Item}
\end{maplelatex}
\begin{maplelatex}\begin{Maple Bullet Item}{
InverseMetric: find the inverse of a metric tensor.}\end{Maple Bullet Item}
\end{maplelatex}
\begin{maplelatex}\begin{Maple Bullet Item}{
KroneckerDelta: find the Kronecker delta tensor of rank \textbf{r}.}\end{Maple Bullet Item}
\end{maplelatex}
\begin{maplelatex}\begin{Maple Bullet Item}{
MetricDensity: use a metric tensor to create a scalar density of a given weight.}\end{Maple Bullet Item}
\end{maplelatex}
\begin{maplelatex}\begin{Maple Bullet Item}{
MultiVector: compute the alternating sum of the tensor product of a list of vector fields.}\end{Maple Bullet Item}
\end{maplelatex}
\begin{maplelatex}\begin{Maple Bullet Item}{
PermutationSymbol: create a permutation symbol.}\end{Maple Bullet Item}
\end{maplelatex}
\begin{maplelatex}
\begin{Maple Bullet Item}{PushPullTensor:  transform a tensor from one manifold or coordinate system to another.}\end{Maple Bullet Item}
\end{maplelatex}
\begin{maplelatex}\begin{Maple Bullet Item}{
RaiseLowerIndices: raise or lower a list of indices of a tensor.}\end{Maple Bullet Item}
\end{maplelatex}
\begin{maplelatex}\begin{Maple Bullet Item}{
RearrangeIndices: rearrange the argument/indices of a tensor.}\end{Maple Bullet Item}
\end{maplelatex}
\begin{maplelatex}\begin{Maple Bullet Item}{
SymmetrizeIndices: symmetrize or skew-symmetrize a list of tensor indices.}\end{Maple Bullet Item}
\end{maplelatex}
\begin{maplelatex}\begin{Maple Bullet Item}{
TensorInnerProduct: compute the inner product of two vectors, forms or tensors with respect to a given metric tensor.}\end{Maple Bullet Item}
\end{maplelatex}

 \textbf{Commands for tensor differentiation}

\begin{maplelatex}\begin{Maple Bullet Item}{
Christoffel: find the Christoffel symbols of the first or second kind for a metric tensor.}\end{Maple Bullet Item}
\end{maplelatex}
\begin{maplelatex}\begin{Maple Bullet Item}{
Connection: define a linear connection on the tangent bundle or on a vector bundle.}\end{Maple Bullet Item}
\end{maplelatex}
\begin{maplelatex}\begin{Maple Bullet Item}{
CovariantDerivative: calculate the covariant derivative of a tensor field with respect to a connection.}\end{Maple Bullet Item}
\end{maplelatex}
\begin{maplelatex}\begin{Maple Bullet Item}{
DirectionalCovariantDerivative: calculate the covariant derivative of a tensor field in the direction of a vector field and with respect to a given connection}\end{Maple Bullet Item}
\end{maplelatex}
\begin{maplelatex}\begin{Maple Bullet Item}{
GeodesicEquations: calculate the geodesic equations for a symmetric linear connection on the tangent bundle \textbf{.}}\end{Maple Bullet Item}
\end{maplelatex}
\begin{maplelatex}\begin{Maple Bullet Item}{
Laplacian: find the Laplacian of a differential form with respect to a metric.}\end{Maple Bullet Item}
\end{maplelatex}
\begin{maplelatex}\begin{Maple Bullet Item}{
ParallelTransportEquations: calculate the parallel transport equations for a linear connection on the tangent bundle or a linear connection on a vector bundle.}\end{Maple Bullet Item}
\end{maplelatex}
\begin{maplelatex}\begin{Maple Bullet Item}{
TensorBrackets: calculate the Schouten bracket and Frolicher-Nijenhuis brackets of tensor fields.}\end{Maple Bullet Item}
\end{maplelatex}
\begin{maplelatex}\begin{Maple Bullet Item}{
TorsionTensor: calculate the torsion tensor for a linear connection on the tangent bundle.}\end{Maple Bullet Item}
\end{maplelatex}

 \textbf{Commands for calculating curvature tensors}

\begin{maplelatex}\begin{Maple Bullet Item}{
Christoffel: find the Christoffel symbols of the first or second kind for a metric tensor.}\end{Maple Bullet Item}
\end{maplelatex}
\begin{maplelatex}\begin{Maple Bullet Item}{
Connection: define a linear connection on the tangent bundle or on a vector bundle.}\end{Maple Bullet Item}
\end{maplelatex}
\begin{maplelatex}\begin{Maple Bullet Item}{
CottonTensor: calculate the Cotton tensor for a metric.}\end{Maple Bullet Item}
\end{maplelatex}
\begin{maplelatex}\begin{Maple Bullet Item}{
CurvatureTensor: calculate the curvature tensor of a linear connection on the tangent bundle or on a vector bundle.}\end{Maple Bullet Item}
\end{maplelatex}
\begin{maplelatex}\begin{Maple Bullet Item}{
EinsteinTensor: calculate the Einstein tensor for a metric.}\end{Maple Bullet Item}
\end{maplelatex}
\begin{maplelatex}\begin{Maple Bullet Item}{
RicciScalar: calculate the Ricci scalar for a metric.}\end{Maple Bullet Item}
\end{maplelatex}
\begin{maplelatex}\begin{Maple Bullet Item}{
RicciTensor: calculate the Ricci tensor of a linear connection on the tangent bundle.}\end{Maple Bullet Item}
\end{maplelatex}
\begin{maplelatex}\begin{Maple Bullet Item}{
RiemannInvariants:  calculate a complete set of scalar curvature invariants in 4 dimensions.}\end{Maple Bullet Item}
\end{maplelatex}
\begin{maplelatex}\begin{Maple Bullet Item}{
SectionalCurvature: calculate the sectional curvature for a metric.}\end{Maple Bullet Item}
\end{maplelatex}
\begin{maplelatex}\begin{Maple Bullet Item}{
WeylTensor: calculate the Weyl curvature tensor of a metric.}\end{Maple Bullet Item}
\end{maplelatex}

\textbf{Commands for calculating special tensor fields}

\begin{maplelatex}\begin{Maple Bullet Item}{
CovariantlyConstantTensors: calculate the covariantly constant tensors with respect to a given metric or connection.}\end{Maple Bullet Item}
\end{maplelatex}
\begin{maplelatex}\begin{Maple Bullet Item}{
KillingSpinors: calculate the Killing spinors for a given spacetime.}\end{Maple Bullet Item}
\end{maplelatex}
\begin{maplelatex}\begin{Maple Bullet Item}{
KillingYanoTensors:  calculate the Killing tensors of a specified rank for a given metric or connection.}\end{Maple Bullet Item}
\end{maplelatex}
\begin{maplelatex}\begin{Maple Bullet Item}{
KillingTensors: calculate the Killing-Yano tensors for a given connection or a given metric.}\end{Maple Bullet Item}
\end{maplelatex}
\begin{maplelatex}\begin{Maple Bullet Item}{
RecurrentTensors: calculate the recurrent tensors with respect to a given metric or connection.}\end{Maple Bullet Item}
\end{maplelatex}

\textbf{Commands for working with Killing tensors}

\begin{maplelatex}\begin{Maple Bullet Item}{
CheckKillingTensor: check that a tensor is the Killing tensor for a metric.}\end{Maple Bullet Item}
\end{maplelatex}
\begin{maplelatex}\begin{Maple Bullet Item}{
IndependentKillingTensors: create a list of linearly independent Killing tensors.}\end{Maple Bullet Item}
\end{maplelatex}
\begin{maplelatex}\begin{Maple Bullet Item}{
KillingBracket: a covariant form of the Schouten bracket for symmetric tensors.}\end{Maple Bullet Item}
\end{maplelatex}
\begin{maplelatex}\begin{Maple Bullet Item}{
SymmetricProductsOfKillingTensors: form all possible symmetric tensors of a given rank.}\end{Maple Bullet Item}
\end{maplelatex}

 \textbf{Commands for the 2-component spinor formalism}

\begin{maplelatex}\begin{Maple Bullet Item}{
BivectorSolderForm:  calculate the rank 4 spin-tensor which maps bivectors to symmetric rank 2 spinors.}\end{Maple Bullet Item}
\end{maplelatex}
\begin{maplelatex}\begin{Maple Bullet Item}{
ConjugateSpinor:  calculate the complex conjugate of a spinor.}\end{Maple Bullet Item}
\end{maplelatex}
\begin{maplelatex}\begin{Maple Bullet Item}{
EpsilonSpinor:  calculate the epsilon spinor in the 2 component spinor formalism.}\end{Maple Bullet Item}
\end{maplelatex}
\begin{maplelatex}\begin{Maple Bullet Item}{
KroneckerDeltaSpinor:  calculate the Kronecker delta spinor in the 2 component spinor formalism.}\end{Maple Bullet Item}
\end{maplelatex}
\begin{maplelatex}\begin{Maple Bullet Item}{
RaiseLowerSpinorIndices:  raise/lower the indices of a spinor or spin-tensor using the epsilon spinors.}\end{Maple Bullet Item}
\end{maplelatex}
\begin{maplelatex}\begin{Maple Bullet Item}{
RicciSpinor:  calculate the rank 4 Ricci spinor corresponding to the trace-free Ricci tensor.}\end{Maple Bullet Item}
\end{maplelatex}
\begin{maplelatex}\begin{Maple Bullet Item}{
SolderForm:  calculate the solder form (or Infeld-van der Waerden symbols) from an orthonormal tetrad.}\end{Maple Bullet Item}
\end{maplelatex}
\begin{maplelatex}\begin{Maple Bullet Item}{
SpinConnection:  calculate the unique spin connection defined by a solder form.}\end{Maple Bullet Item}
\end{maplelatex}
\begin{maplelatex}\begin{Maple Bullet Item}{
SpinorInnerProduct:   contract all spinor indices of 2 spin-tensors using the epsilon spinors.}\end{Maple Bullet Item}
\end{maplelatex}
\begin{maplelatex}\begin{Maple Bullet Item}{
WeylSpinor:  calculate the rank 4 Weyl spinor corresponding to the Weyl tensor.}\end{Maple Bullet Item}
\end{maplelatex}
 \textbf{Commands for the Newman-Penrose formalism}

\begin{maplelatex}\begin{Maple Bullet Item}{
GRQuery:  verify various properties of spacetimes.}\end{Maple Bullet Item}
\end{maplelatex}
\begin{maplelatex}\begin{Maple Bullet Item}{
NPBianchiIdentities:  calculate the Bianchi identities in the Newman-Penrose formalism.}\end{Maple Bullet Item}
\end{maplelatex}
\begin{maplelatex}\begin{Maple Bullet Item}{
NPCurvatureScalars:  calculate the Ricci scalars and the Weyl scalars in the Newman-Penrose formalism.}\end{Maple Bullet Item}
\end{maplelatex}
\begin{maplelatex}\begin{Maple Bullet Item}{
NPDirectionalDerivatives:  define the directional derivative operators used in the Newman-Penrose formalism.}\end{Maple Bullet Item}
\end{maplelatex}
\begin{maplelatex}\begin{Maple Bullet Item}{
NPRicciIdentities:  calculate the Ricci identities in the Newman-Penrose formalism.}\end{Maple Bullet Item}
\end{maplelatex}
\begin{maplelatex}\begin{Maple Bullet Item}{
NPSpinCoefficients:  calculate the Newman-Penrose spin coefficients.}\end{Maple Bullet Item}
\end{maplelatex}
\begin{maplelatex}\begin{Maple Bullet Item}{
NullTetrad:  calculate a null tetrad from an orthonormal tetrad.}\end{Maple Bullet Item}
\end{maplelatex}
\begin{maplelatex}\begin{Maple Bullet Item}{
NullTetradTransformation:  apply a Lorentz transformation to a null tetrad.}\end{Maple Bullet Item}
\end{maplelatex}
\begin{maplelatex}\begin{Maple Bullet Item}{
OrthonormalTetrad:  calculate an orthonormal tetrad from a null tetrad.}\end{Maple Bullet Item}
\end{maplelatex}
\begin{maplelatex}\begin{Maple Bullet Item}{
SpacetimeConventions:  a description of the tensor and spinor conventions used by the DifferentialGeometry package.}\end{Maple Bullet Item}
\end{maplelatex}

 \textbf{Commands for the algebraic classification of spacetimes}

\begin{maplelatex}\begin{Maple Bullet Item}{
IsotropyType:  determine the isotropy type of the isotropy subalgebra of infinitesimal isometries.}\end{Maple Bullet Item}
\end{maplelatex}
\begin{maplelatex}\begin{Maple Bullet Item}{
PetrovType:  determine the Petrov type of a spacetime from the Weyl tensor.}\end{Maple Bullet Item}
\end{maplelatex}
\begin{maplelatex}\begin{Maple Bullet Item}{
SegreType:  determine the Plebanski-Petrov type and the Segre type of a trace-free, rank 2 symmetric tensor.}\end{Maple Bullet Item}
\end{maplelatex}
\begin{maplelatex}\begin{Maple Bullet Item}{
SubspaceType:  determine the signature of the metric restricted to a subspace.}\end{Maple Bullet Item}
\end{maplelatex}

 \textbf{Commands for calculation of tensorial invariants}

\begin{maplelatex}\begin{Maple Bullet Item}{
KillingBracket:  a covariant form of the Schouten bracket for symmetric tensors.}\end{Maple Bullet Item}
\end{maplelatex}
\begin{maplelatex}\begin{Maple Bullet Item}{
KillingSpinors:  calculate the Killing spinors for a given spacetime.}\end{Maple Bullet Item}
\end{maplelatex}
\begin{maplelatex}\begin{Maple Bullet Item}{
KillingTensors:  calculate the Killing tensors of a specified rank for a given metric or connection.}\end{Maple Bullet Item}
\end{maplelatex}
\begin{maplelatex}\begin{Maple Bullet Item}{
KillingVectors:  calculate the Killing vectors for a given metric.}\end{Maple Bullet Item}
\end{maplelatex}
\begin{maplelatex}\begin{Maple Bullet Item}{
KillingYanoTensors:  calculate the Killing-Yano tensors for a given connection or a given metric.}\end{Maple Bullet Item}
\end{maplelatex}

 \textbf{Commands for field theory}

\begin{maplelatex}\begin{Maple Bullet Item}{
BelRobinson:  calculate the rank 4 Bel-Robinson tensor for a metric.}\end{Maple Bullet Item}
\end{maplelatex}

\begin{maplelatex}\begin{Maple Bullet Item}{
DivergenceIdentities:  check the divergence identity for various energy-momentum tensors.}\end{Maple Bullet Item}
\end{maplelatex}
\begin{maplelatex}\begin{Maple Bullet Item}{
EnergyMomentumTensor:  calculate the energy-momentum tensor for various fields (scalar, electromagnetic, dust, ...).}\end{Maple Bullet Item}
\end{maplelatex}
\begin{maplelatex}\begin{Maple Bullet Item}{
MatterFieldEquations:  calculate the field equations for various field theories (scalar, electromagnetic, dust, ...).}\end{Maple Bullet Item}
\end{maplelatex}

 \end{document}